\documentclass[aps,prl,twocolumn,preprintnumbers,amsmath,amssymb]{revtex4-2}
\usepackage{graphicx}
\usepackage{dcolumn}
\usepackage{bm}
\usepackage{subfigure}
\usepackage{multirow}
\usepackage{color}
\newcommand{\thetap}{\theta_{\rm p}}
\newcommand{\rhoe}{\rho_{\rm e}}
\newcommand{\te}{t_{\rm e}}
\newcommand{\PB}{P_B}

\definecolor{blue}{rgb}{0,0,1}
\definecolor{red}{rgb}{1,0,0}
\definecolor{green}{rgb}{0,1,0}

\begin{document}

\title{Universal Critical Behavior of Percolation in Orientationally Ordered Janus Particles \\ and Other Anisotropic Systems}
\author{Hao Hu}
\email{huhao@ahu.edu.cn}
\affiliation{School of Physics and Optoelectronic Engineering, Anhui University, Hefei 230601, China}
\author{Robert M. Ziff}
\affiliation{Center for the Study of Complex Systems and Department of Chemical Engineering, University of Michigan, Ann Arbor, Michigan 48109-2800, USA}
\author{Youjin Deng}
\affiliation{Department of Modern Physics, University of Science and Technology of China, Hefei 230026, China}
\affiliation{MinJiang Collaborative Center for Theoretical Physics, College of Physics and Electronic Information Engineering, Minjiang University, Fuzhou 350108, China}

\begin{abstract}
        We combine percolation theory and Monte Carlo simulation to study in two dimensions the connectivity of 
	an equilibrium lattice model of interacting Janus disks which self-assemble into an orientationally 
	ordered stripe phase at low temperature. As the patch size is increased or the temperature is lowered, 
	clusters of patch-connected disks grow, and a percolating cluster emerges at a threshold.
	In the stripe phase, the critical clusters extend longer in the direction parallel to the stripes
	than in the perpendicular direction, and percolation is thus anisotropic.
	It is found that the critical behavior of percolation in the Janus system is consistent with that of 
	standard isotropic percolation, when an appropriate spatial rescaling is made.
	The rescaling procedure can be applied to understand other anisotropic systems, 
	such as the percolation of aligned rigid rods and of the $q$-state Potts model with anisotropic interactions. 
\end{abstract}
\maketitle

The novelty of Janus particles~\cite{Casagrande1988}, which have two surface areas of different properties, 
 was appreciated very early in the field of soft matter~\cite{deGennes1992}. 
 Nowadays the particles can be synthesized by various methods
 and are used as surfactants, micromotors, displays, catalysts, 
 biosensors, etc.~\cite{Lahann2011,Zhang2017,Kirillova2019,Zhang2021}.
 The possibilities offered by Janus particles mainly depend on the fact that heterogeneous surfaces 
 lead to anisotropic interactions.
 Harnessing the unique interactions, unconventional self-assembled structures can be made, such as gases of micelles~\cite{Sciortino2009,Iwashita2013}, entropy stabilized 
 open crystals~\cite{Chen2011, Mao2013}, and crystals of orientational 
 order~\cite{Sciortino2009,Sciortino2010,Vissers2013,Preisler2013,Villegas2014,Preisler2014,Iwashita2017,Liang2021,Shin2014,Iwashita2014,Jiang2014,Rezvantalab2016,Iwashita2016,Mitsumoto18,Huang2019,Huang2020}.
 However, in orientationally ordered Janus systems, long-range connectivity behavior, 
 which is important to understand transport~\cite{Balberg2017}, mechanical~\cite{Tanaka2019}, dynamical~\cite{Cho2020,Falsi2021}, and other properties~\cite{Stauffer1992,Sahimi1994}, remains
 largely unexplored~\cite{Sciortino2010,Iwashita2014,Wang2022}.

 Recently it was found that, in the orientationally ordered nematic phase of slender nanoparticles,
 the coupling between the particle density and the orientational order gives rise to
 interesting nonmonotonic behavior of the percolation threshold as a function of the density~\cite{Finner2019a,Finner2019b}. 
 Percolation deals with long-range connectivity and is one of the most applied models in statistical physics~\cite{Stauffer1992,Sahimi1994,Araujo2014,Saberi2015}. 
 At the percolation threshold, a system-spanning connected cluster first appears,
 and there exist various universal properties, e.g., critical exponents, dimensionless quantities, correlation and scaling functions~\cite{Pelissetto2002}.
 For percolation in anisotropic systems, {while} some results 
 have been obtained on thresholds and critical exponents~\cite{Sykes1963,SykesEssam1964,Balberg1985,Kenyon2004,GM2014,Tarasevich12,Longone2012,Longone19,Finner2019a,Finner2019b,Falsi2021}, 
 other universal critical properties, such as the continuous change
 of dimensionless wrapping probabilities of aligned rigid rods~\cite{Tarasevich12,Longone2012,Longone19}, are not well understood.
 
 For most systems of Janus particles, positional and rotational motions are coupled,
 which leads to complex phase behavior. However, for close-packed crystals of
 Janus particles, positional vibrations are much less important. 
 Thus close-packed Janus systems provide a platform where one can 
 tune controlling parameters, e.g., the temperature or pressure, to explore 
 orientaional order driven by rotational fluctuations
 ~\cite{Jiang2014,Shin2014,Iwashita2014,Iwashita2016,Rezvantalab2016,Mitsumoto18,Huang2019,Patrykiejew19,Patrykiejew20}.
 In this Letter, by exploring anisotropic interacting close-packed Janus disks in two dimensions (2D), 
 we ask how thermal rotational fluctuations affect universal critical behavior of anisotropic percolation
 in crystal phases of orientational orders, and how the results can be generalized to understand percolation in other anisotropic systems.

 We use a simple model of Janus disks on the triangular lattice, 
 where particles interact with the Kern-Frenkel potential~\cite{Kern2003}.
 Combining percolation theory and Monte Carlo (MC) simulation, we find that,
 in the orientationally ordered stripe phase~\cite{Mitsumoto18}, though critical exponents are consistent 
 with standard isotropic percolation, universal values of dimensionless quantities (e.g., Binder-like ratios
 and wrapping probabilities~\cite{Hu2015}) change continuously along the percolation line. Using theoretical results 
 for wrapping probabilities of standard percolation in 2D~\cite{diFrancesco87,Pinson94,Ziff99,supMat},
 we find an effective aspect ratio $\rhoe$ to perform a spatial rescaling and relate quantitatively
 universal critical behavior of percolation 
 in the stripe phase to that of standard percolation. Thus the mechanism underlying 
 the continuous variations of dimensionless quantities is that anisotropic interactions 
 in the stripe phase cause connectivity correlations to be anisotropic, but the behavior 
 can be captured by standard percolation. 
 We then show that the mechanism can also explain anisotropic percolation behavior 
 in other systems, such as systems of particles with anisotropic shapes (e.g., aligned rigid rods~\cite{Tarasevich12,Longone2012, Longone19}),
 and systems of anisotropic bond-placing rules.
 For the latter systems, by studying anisotropic bond percolation~\cite{Sykes1963,SykesEssam1964} on the triangular lattice, we demonstrate 
 by the isoradial-graph method~\cite{Kenyon2004,GM2014} that a more general relation between 
 anisotropic percolation and standard percolation 
 requires an effective shear transformation involving both $\rhoe$ and an effective boundary twist $t_{\rm e}$. 
 In terms of conformal invariance and universality, statistical models on isoradial graphs have nice properties~\cite{Copin2018},
 which we further use to derive {$\rhoe$ and $\te$ for understanding} universal critical behavior of percolation in the anisotropic $q$-state Potts models~\cite{supMat}.
 The results may help design materials with tunable anisotropic connectivity-related 
 properties such as ferroelectricity~\cite{Falsi2021}, conductivity~\cite{Ledezma2011,Ackermann2016}, and photovoltaics~\cite{Thorkelsson2015}.

\begin{figure}[htbp] 
   \centering
   \includegraphics[width=3.3 in]{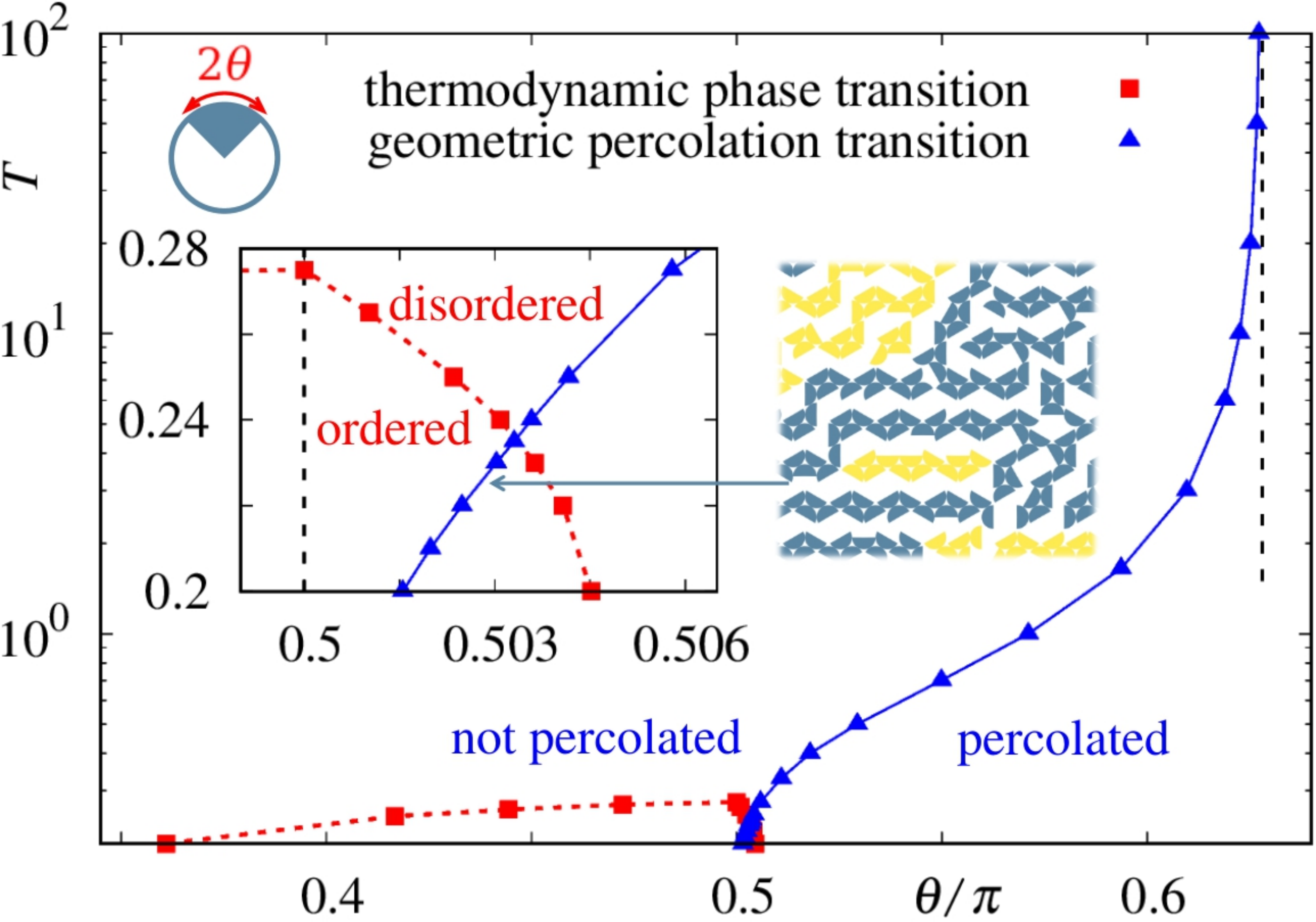} 
   \caption{Phase diagram of close-packed Janus disks in 2D. 
        The system is ordered below the phase-transition line with squares, 
	and it is percolated to the right side of the percolation-transition line with triangles.
	According to whether the system is ordered or percolated, 
	the diagram is divided into four regions: an ordered and (not) percolated region,
	and a disordered and (not) percolated region.
	Inset: enlargement near $\theta/\pi=0.503$,
	as well as a snapshot of an ordered and percolated configuration, with the largest cluster being dark blue.
	Vertical dashed lines indicate the position of $\theta/\pi=1/2$ 
	and the infinite-temperature percolation threshold $\thetap/\pi \simeq 0.628$~\cite{Wang2022}. 
	Lines going through data points are added to guide the eye, 
	and the error bars are smaller than or comparable with the symbols.}
   \label{fig:phase_diagram}
\end{figure}

The model system consists of close-packed Janus disks with monodisperse patch sizes in 2D.
Rhombus-shaped $L \times L$ triangular lattices with periodic boundary conditions are used, 
and each lattice site is occupied by a disk with diameter $1$. 
To allow only rotational motions, the center of each disk is fixed at a lattice site.
As shown at top left of Fig.~\ref{fig:phase_diagram}, the dark sector represents a patch on the disk 
which spans an angle of $2\theta$. The half-patch angle $\theta$ (i.e., the Janus balance~\cite{Jiang2008}) characterizes the patch size.
The Janus disks interact with a Kern-Frenkel potential~\cite{Kern2003}: two nearest-neighbor disks contribute 
an energy $-\epsilon$ if the two patches on them cover the same edge and touch each other; 
otherwise, they contribute a zero energy. When studying percolation,
two disks are regarded as connected when they interact with an energy $-\epsilon$.
The unit of temperature $T$ is $\epsilon/k_{\rm B}$, where $k_{\rm B}$ is the Boltzmann constant.

Figure~\ref{fig:phase_diagram} shows the phase diagram in the $T-\theta$ plane.
Previously it was found that, for $1/3 < \theta/\pi \le 1/2$, there is a continuous thermodynamic phase transition 
from a high-$T$ disordered phase to a low-$T$ orientationally ordered stripe phase~\cite{Mitsumoto18}. 
For close-packed Janus particles in 2D continuum space, 
preliminary results showed that this thermodynamic phase transition is still continuous~\cite{Liang2021}.
Since at $T=\infty$ the percolation threshold 
is $\thetap/\pi=0.627\,765\,41(3)$~\cite{Wang2022}, to explore connectivity at finite $T$,
we need first understand thermodynamic behavior for $\theta/\pi > 1/2$.
We performed extensive simulation using the Metropolis algorithm, 
{where in a MC sweep the disks are sequentially visited and independently proposed to rotate 
by a random angle in the range $[-\pi,\pi)$.}
An orientational order parameter based on the structure factor and the associated Binder ratio 
defined in Ref.~\cite{Mitsumoto18} were sampled~\cite{supMat}.
We find that the continuous phase transition from the disordered phase to the stripe phase 
also exists for $\theta/\pi>1/2$, and determine the phase transition line~\cite{supMat}
as plotted in Fig.~\ref{fig:phase_diagram}.

To explore connectivity of the Janus disks,
we combine the critical polynomial method~\cite{Scullard2008,Scullard2010,
Scullard2011,Scullard2012,MertensZiff2016,Scullard2020} with MC simulation.
In the probabilistic geometric interpretation~\cite{Scullard2012}, for standard percolation
in 2D, the critical polynomial is defined as $\PB \equiv R_2-R_0$, where $R_2$ is 
the probability that there exists a cross-wrapping cluster 
and $R_0$ is the probability of no wrapping.
The root of $\PB=0$ gives the percolation threshold when $L \rightarrow \infty$. 
The critical polynomial has been demonstrated to be very powerful 
in determining percolation thresholds in 2D~\cite{Scullard2020,Xu2021,Wang2022}.
Wrapping probabilities were sampled in our simulation.
Near the whole percolation line, 
curves of $\PB(\theta,L)$ cross, and the crossing points converge quickly 
to $\PB=0$. The fact that  $\PB(\thetap,L\rightarrow\infty)=0$
suggests that the percolation transition belongs to the universality class
of standard percolation, since this limit is not true for other models such as the $q$-state Potts model~\cite{Jacobsen2013}. 
We perform finite-size scaling analysis and find that critical exponents
indeed take values for standard percolation~\cite{supMat}.
The estimated percolation thresholds at different $T$ 
are shown by blue triangles in Fig.~\ref{fig:phase_diagram}.
It is seen that, as $T$ drops, the value of $\thetap$ 
decreases and approaches $\theta = \pi/2$ in the low-$T$ limit. 
Thus, the orientationally ordered stripe phase is not guaranteed to be percolated. 

\begin{figure}[htbp] 
   \centering
   \includegraphics[width=1.7 in]{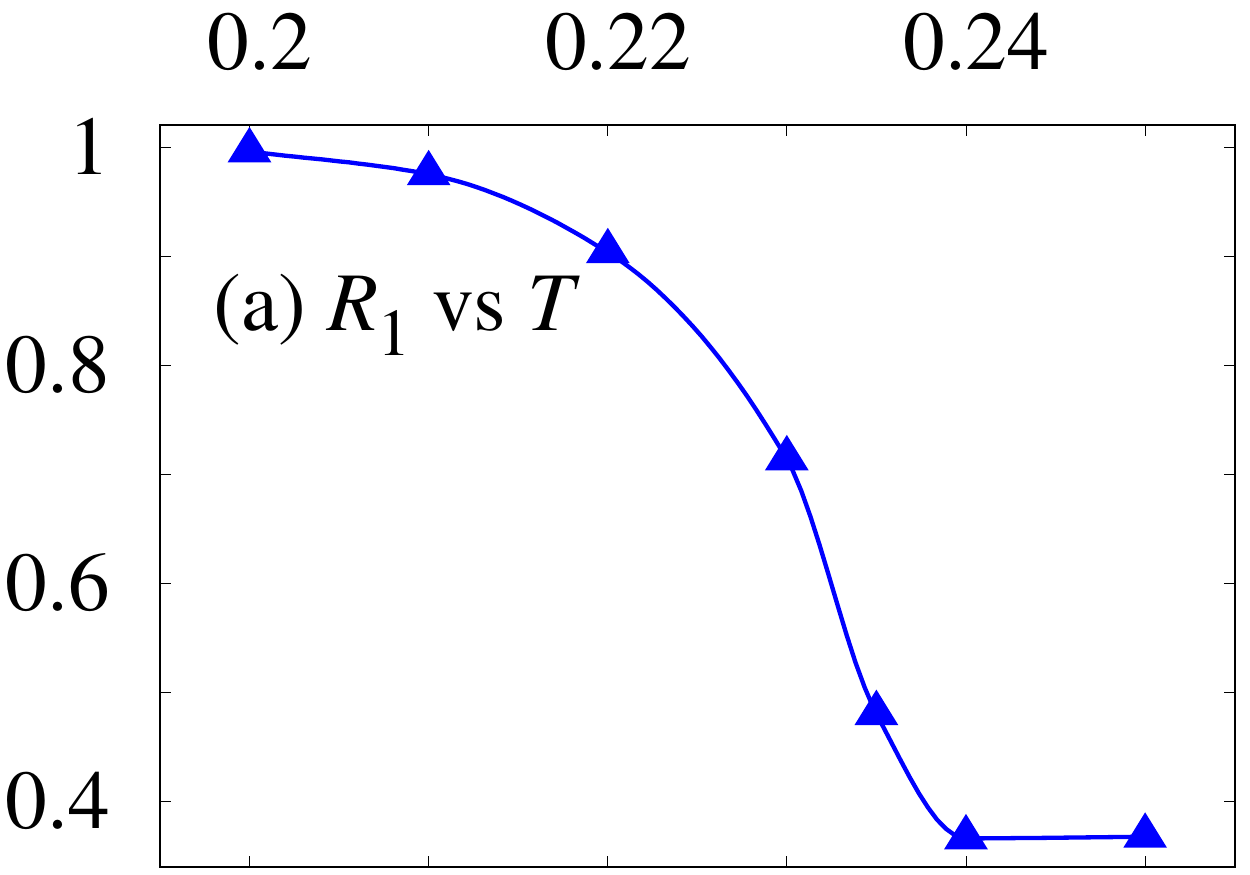} 
	\hspace{-2mm}\includegraphics[width=1.7 in]{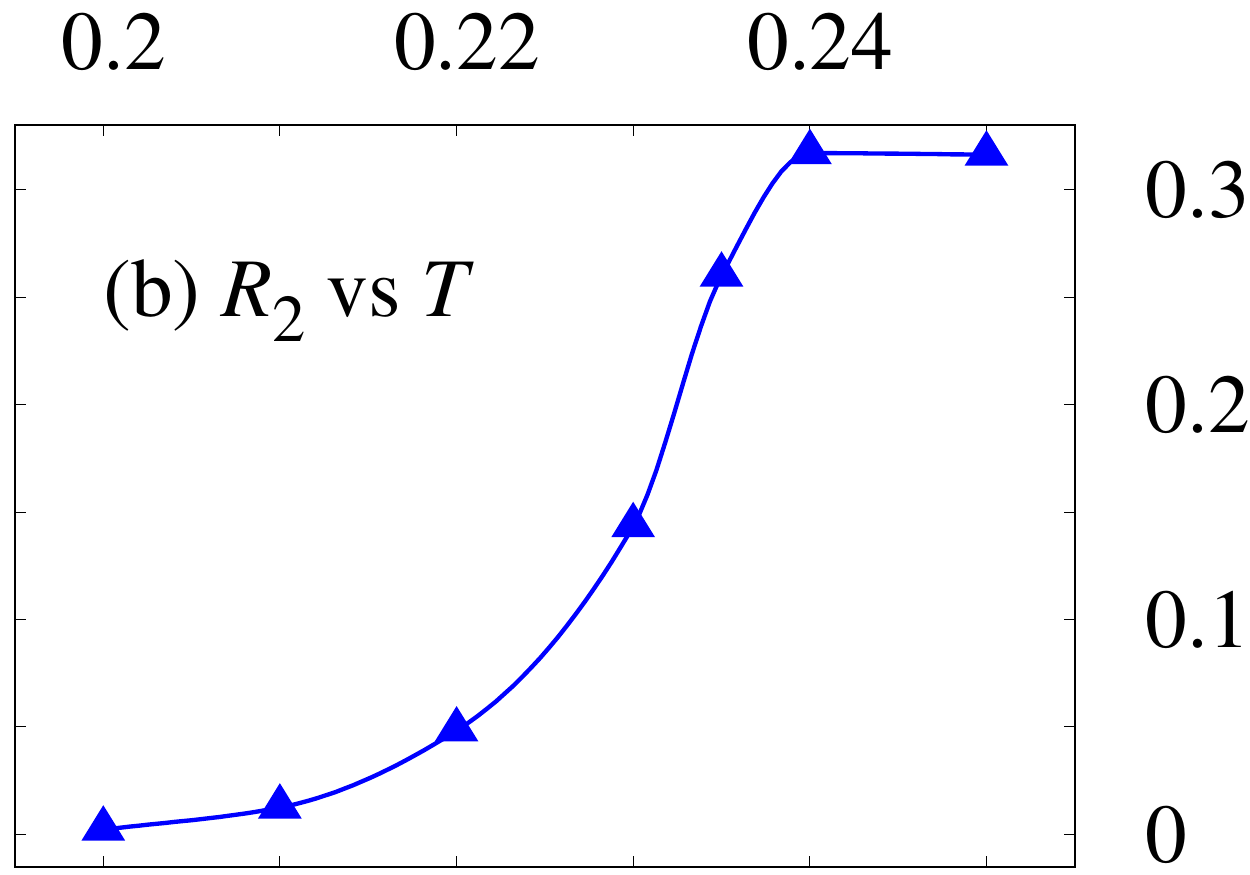} \\ 
	\vspace{-0.8mm}
   \includegraphics[width=1.7 in]{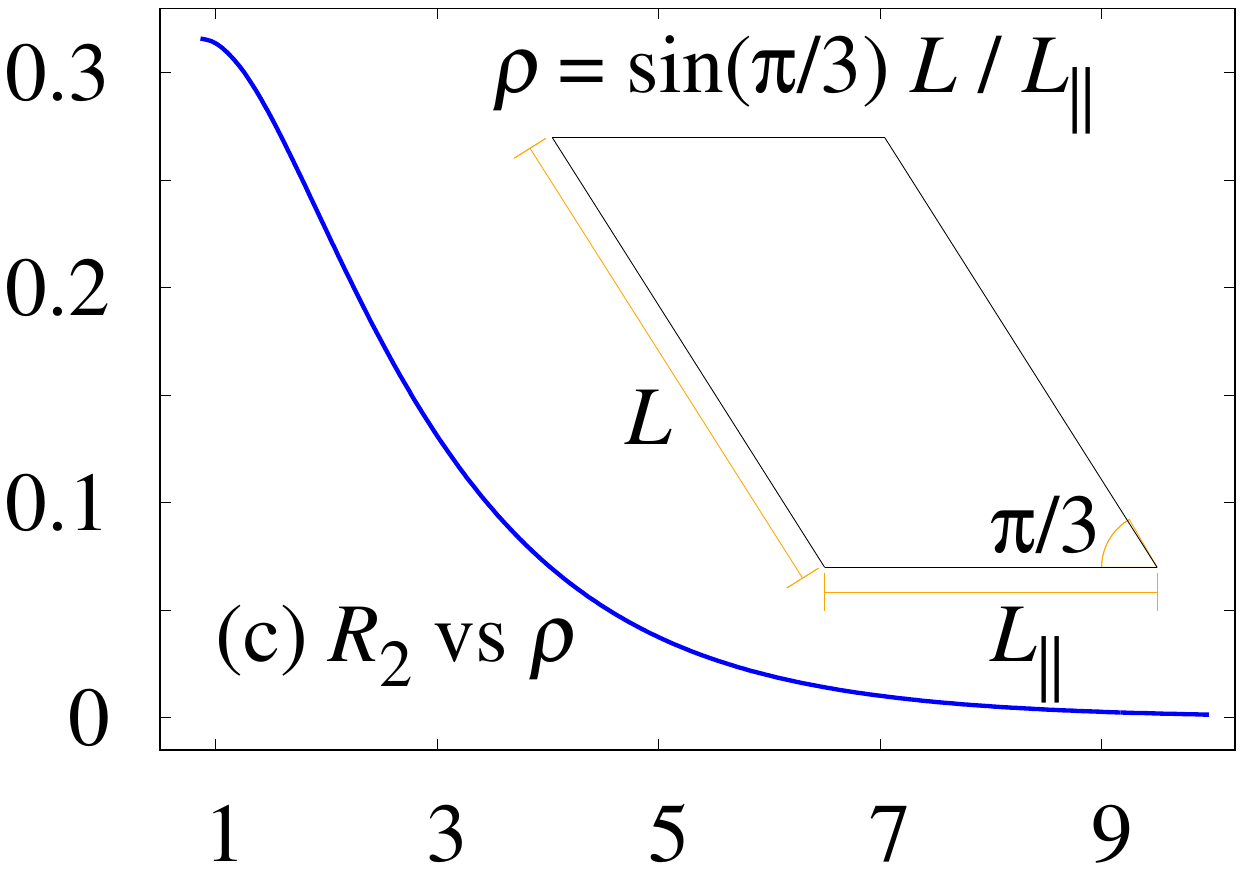} 
	\hspace{-2mm}\includegraphics[width=1.7 in]{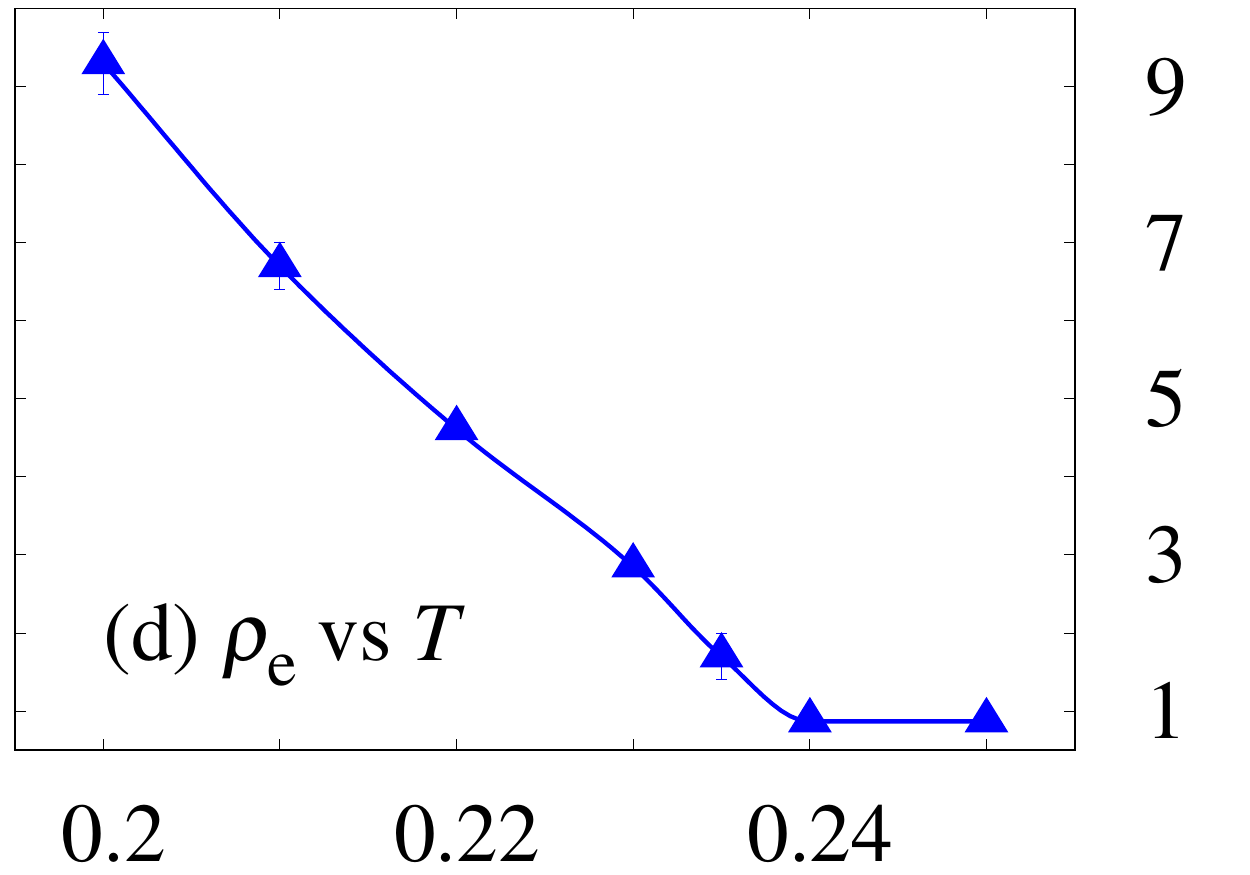}  
	\caption{Critical wrapping probabilities along the percolation line 
	{[i.e., solid curve with triangular points $T(\theta)$ in Fig.~\ref{fig:phase_diagram}]}
	of the Janus system. (a) Probability of wrapping in only one direction $R_1$ vs $T$.
	(b) Wrapping probability $R_2$ vs $T$.
	(c) Theoretical curve of $R_2$ vs the aspect ratio $\rho$
	for standard percolation on parallelogram-shaped periodic lattices.
	(d) The effective aspect ratio $\rhoe$ vs $T$ for the Janus system.} 
   \label{fig:R_ra_vs_T}
\end{figure}

From the theory of critical phenomena, scale invariance at the percolation threshold is related to 
the fact that many dimensionless quantities are independent of the system size, 
if finite-size corrections are neglected. Critical values of dimensionless quantities are ``universal"~\cite{Pelissetto2002}
in the sense that a quantity holds the same value for different lattices, short-range interactions, etc. 
For wrapping probabilities, along the percolation line 
we find that, in the disordered phase, they indeed take the same values
as those for standard percolation for systems of the same shape~\cite{Langlands1992,Pinson94}.
However, in the stripe phase ($T < 0.24$), we find that
they change continuously as shown in Figs.~\ref{fig:R_ra_vs_T}(a) and ~\ref{fig:R_ra_vs_T}(b).
This implies that the orientational order affects ``universal" critical properties.

\begin{figure}[htbp] 
   \centering
   \includegraphics[width=3.4 in]{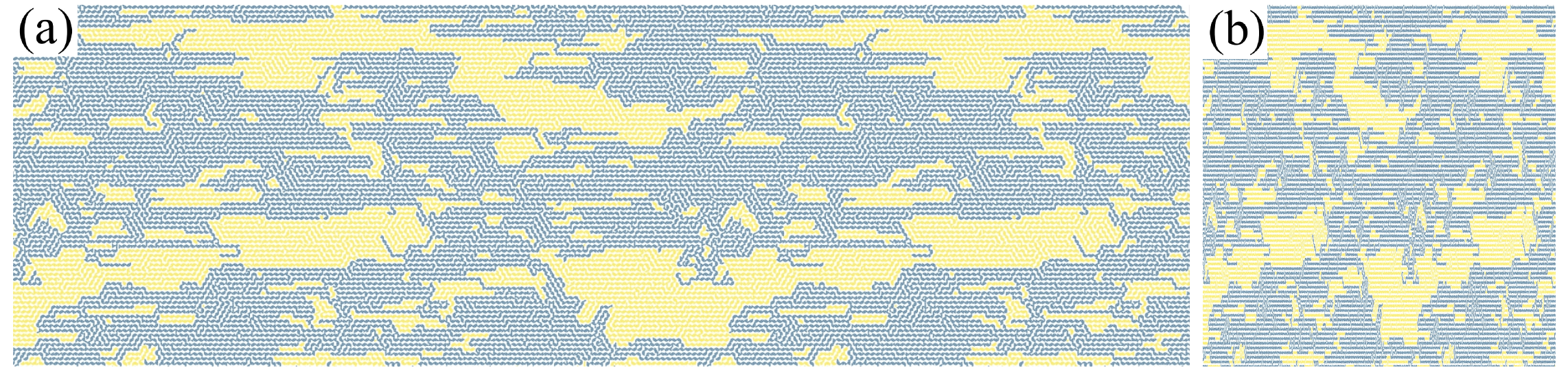} \\ 
   \includegraphics[height=1.35 in]{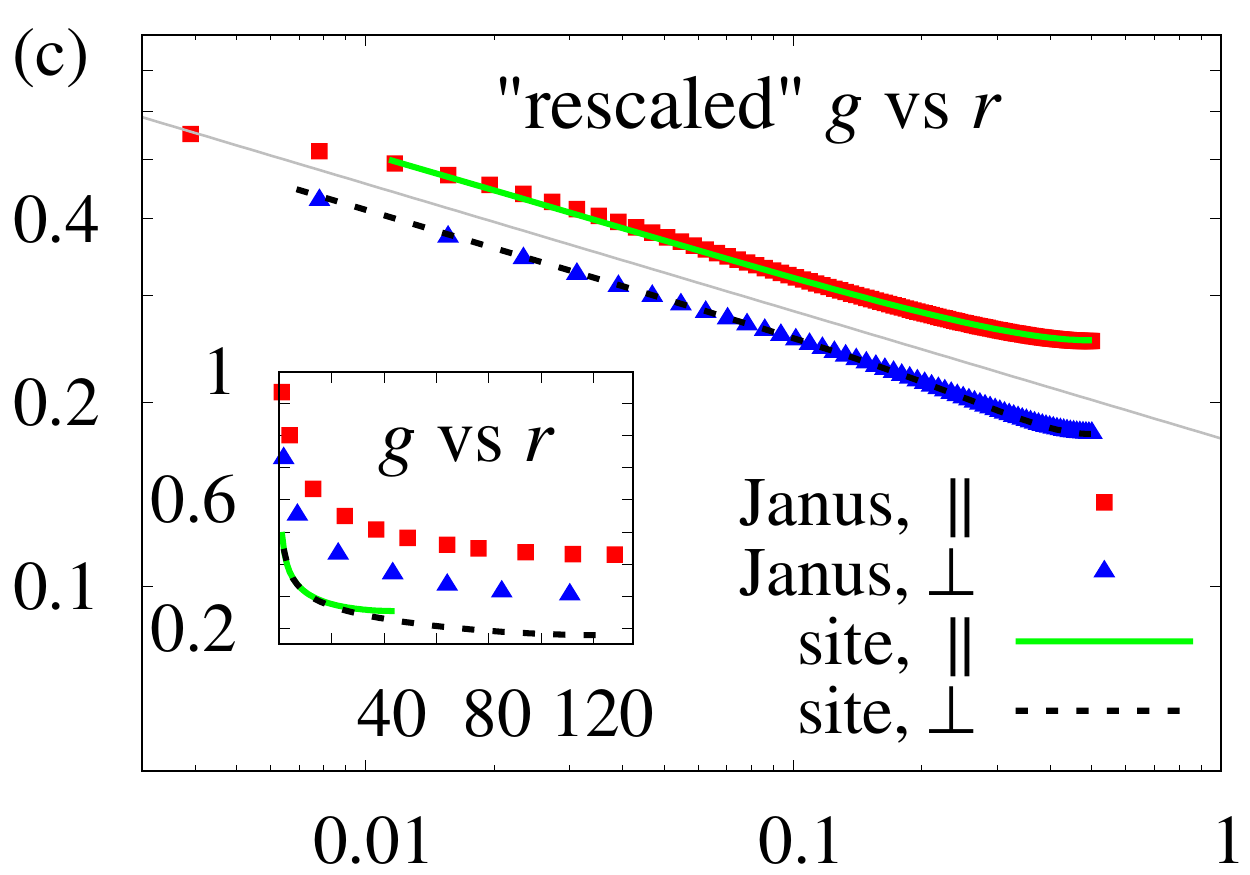} 
   \includegraphics[height=1.35 in]{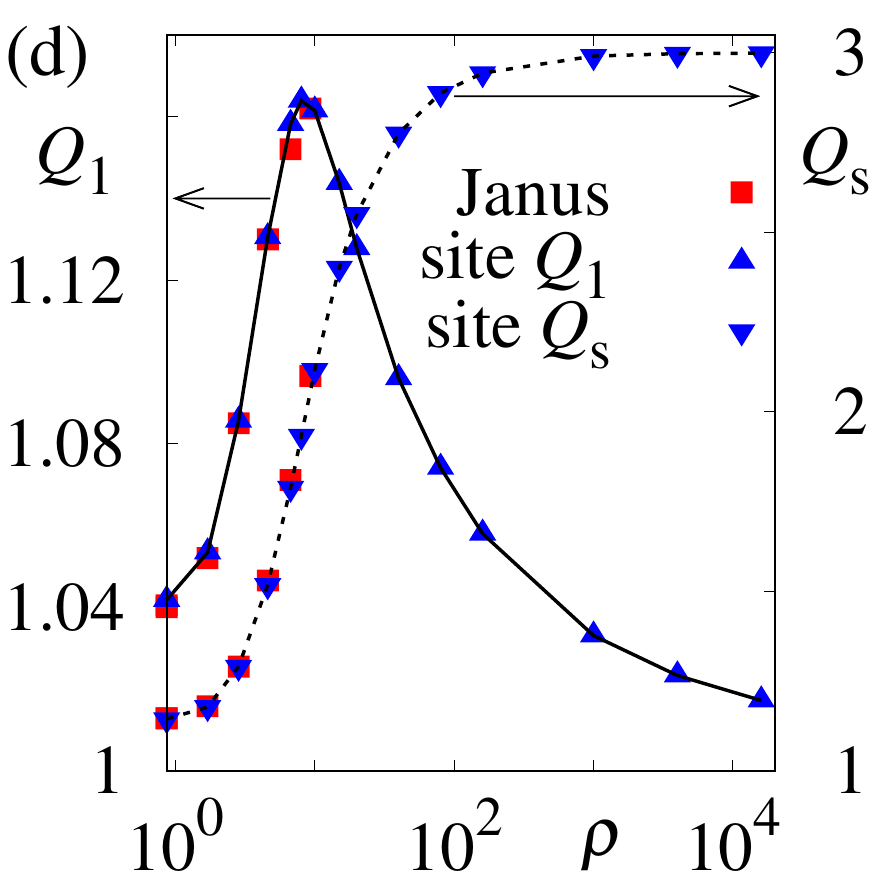} 
	\caption{Relations between percolation in the stripe phase of the Janus system 
  	         and standard site percolation on the triangular lattice.
		 (a) A snapshot of the Janus system
		 at $\thetap/\pi=0.503\,027(2)$, with $T=0.23$ and $L=256$,
		 which is ordered in the parallel (horizontal) direction. 
		 The dark blue region represents the largest wrapping cluster, 
		 with its holes being light yellow. 
		 (b) Isotropic configuration after rescaling the parallel 
		 direction by $\sqrt{3}/(2\rhoe)$, with $\rhoe\simeq2.85$.
		 (c) Collapse of critical correlations of the Janus system 
		 and those of site percolation on the triangular lattice 
		 with size $L_{\parallel} \times L = 88 \times 290$.
		 The light gray line has a slope $-5/24$ from percolation universality.
		 (d) Collapse of critical ratios $Q_1$ and $Q_s$~\cite{Deng2005,Hu2012} 
		 of the Janus system and those of site percolation with $\rho=\rhoe$.} 
   \label{fig:relations}
\end{figure}

It has been known that ``universal" values of critical dimensionless quantities still depend on
factors such as the system shape and boundary conditions~\cite{Langlands1992,Kamieniarz1993,Pinson94,Ziff99,Malakis2014},
anisotropy of couplings~\cite{Chen2004,Selke2005,Selke2009,Kastening2013,Hobrecht2019,Dohm2019,Dohm2021,Dohm2021b}, and statistical ensembles~\cite{Hu2015}.
Thus we are interested in whether or how our results above could be connected with existing results 
for standard percolation. From the configurations near $\thetap$
in the stripe phase, it is seen that clusters are longer in the parallel direction 
than in the perpendicular direction, as exemplified 
in Fig.~\ref{fig:relations}(a). After rescaling the parallel direction by an appropriate factor, 
the configuration becomes isotropic, as shown in Fig.~\ref{fig:relations}(b). 
The rescaling transforms an anisotropic system with size $L \times L$ to an effective isotropic
system with size $L_{\parallel} \times L$ where $L_{\parallel}<L$. 
This leads us to hypothesize that
anisotropic percolation of the $L \times L$ rhombus-shaped Janus system is related to 
standard percolation on a parallelogram-shaped triangular lattice of size $L_{\parallel} \times L$ as depicted in the inset of Fig.~\ref{fig:R_ra_vs_T}(c), 
i.e., the Janus system has an effective aspect ratio $\rhoe = \rho \equiv L_{\perp}/L_{\parallel} = \sin(\pi/3) L / L_{\parallel}$, 
and the rescaling factor is $L_{\parallel}/L = \sqrt{3} / (2\rhoe)$. 

For standard percolation in 2D, 
values of critical wrapping probabilities depend only on $\rho$ and 
the boundary twist $t$ (being $\rho/\sqrt{3}$ for the parallelogram-shaped periodic 
triangular lattice), and their exact expressions
are available~\cite{diFrancesco87,Pinson94,Ziff99,supMat}, 
as illustrated for $R_2$ in Fig.~\ref{fig:R_ra_vs_T}(c).
Thus one can compare numerical values of critical wrapping probabilities 
with the theoretical values to get the $\rhoe$ value.
In this way, we get the dependence of $\rhoe$ on $T$ for the Janus system, 
as shown in Fig.~\ref{fig:R_ra_vs_T}(d).
While $\rhoe$ equals $\sqrt{3}/2$ in the disordered phase,
it monotonically increases as $T$ drops in the stripe phase,
and should approach infinity in the low-$T$ limit. 

With the obtained $\rhoe$, we test our above hypothesis as follows. 
Beside the direct view from Fig.~\ref{fig:relations}(a) 
to Fig.~\ref{fig:relations}(b), we calculate the correlations 
$g(r)$ (probabilities of two particles at a distance $r$ in the same cluster)
in both the parallel and perpendicular directions at $\thetap$.
As shown in Fig.~\ref{fig:relations}(c) for $T=0.23$, the correlations of the Janus system 
can be collapsed into those of standard site percolation by (1) defining the $x$ axis
to be $r_{\parallel}/L_{\parallel} = r/L$ and $r_{\perp }/L_{\perp} = 2 r/ \sqrt{3} L$,
thus $\tilde{g}_{\parallel}(r/L)=g_{\parallel}(r)$ 
and $\tilde{g}_{\perp}(2 r / \sqrt{3}L)=g_{\perp}(r)$, and (2) multiplying
$\tilde{g}$ of the Janus system
by a nonuniversal constant (being $0.588$ for $T=0.23$)~\cite{supMat}.
Further, we compare critical values of two dimensionless ratios
at $\rho=\rhoe$. They are defined as~\cite{Deng2005,Hu2012}
$Q_1 = \langle C_1^2 \rangle / \langle C_1 \rangle ^2$,
$Q_s = \langle 3 S_2^2 - 2 S_4 \rangle / \langle S_2 \rangle ^2$, 
where $C_1$ ($C_i$ for $i \ne 1$) is the size of the largest cluster (other clusters)
and $S_l = \sum_i C_i^l$ is the $l$th moment of cluster sizes.
It can be seen from Fig.~\ref{fig:relations}(d) that
critical values of the two ratios for the Janus system are consistent 
with those of standard site percolation. 
Thus the Janus system is quantitatively related to standard percolation through $\rhoe$. 

\begin{figure}[htbp] 
   \centering
	\includegraphics[width=1.74 in]{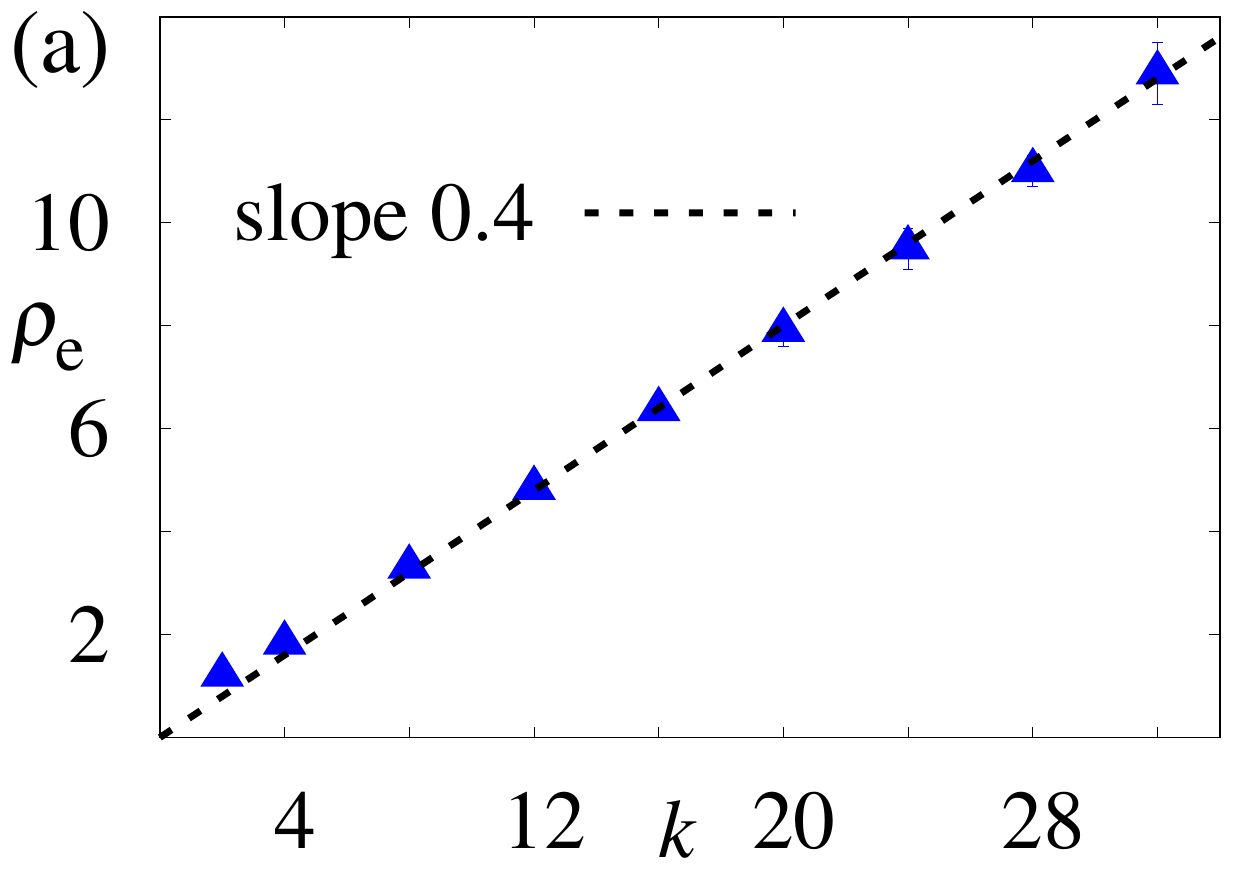}
	\hspace{1mm}
	\includegraphics[width=1.54 in]{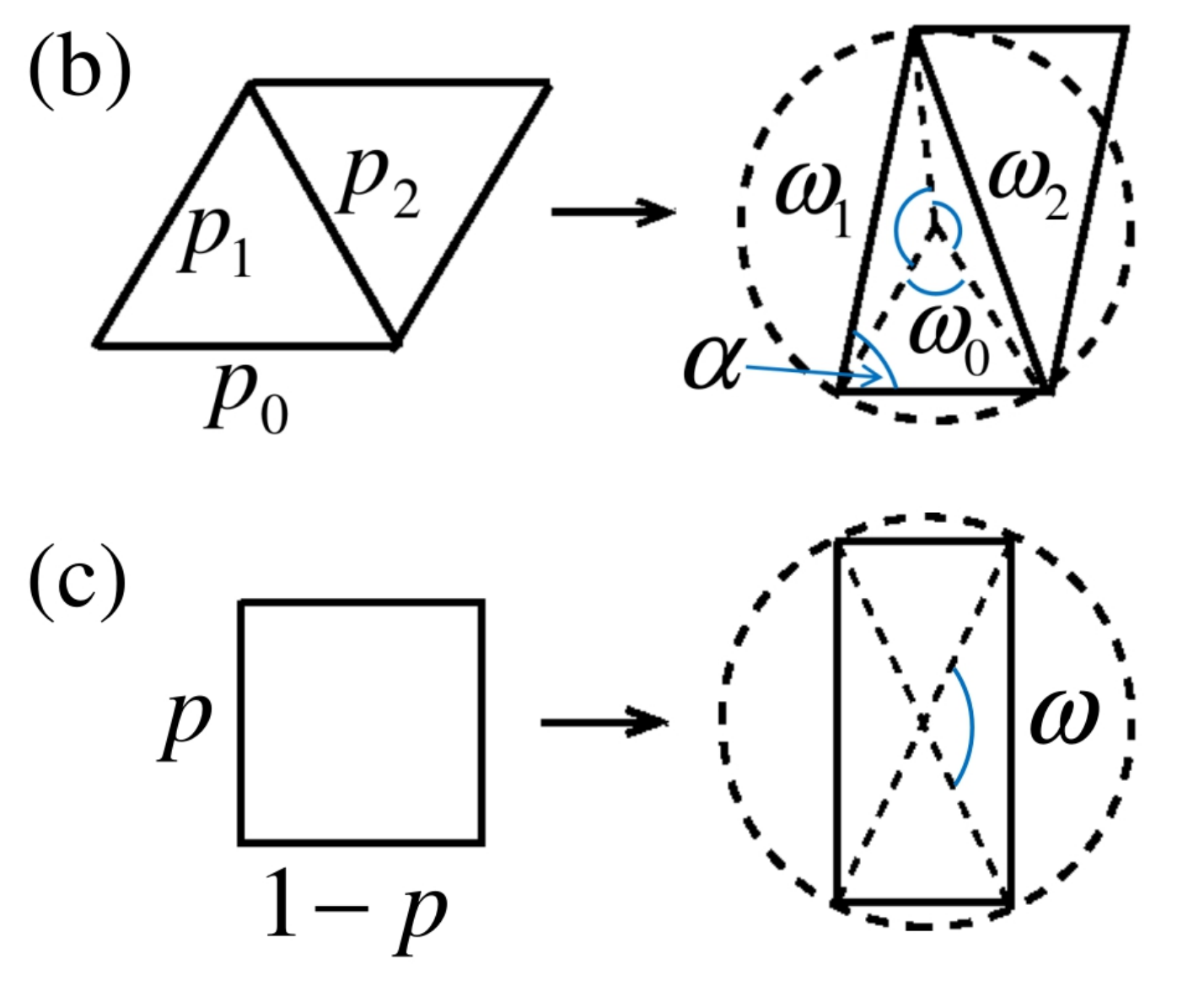} 
	\caption{
		(a) $\rhoe$ vs $k$ for percolation of aligned rigid rods.
		(b), (c) Sketch of the isoradial mapping for anisotropic bond percolation
		on triangular {and square lattices}.
		}
   \label{fig:kmers-isoradial}
\end{figure}

While the anisotropy arises from the emergent orientational order 
in the stripe phase, it can also come from anisotropic constituent 
particles or anisotropic bond-placing rules.
For the former, we consider aligned rigid rods of various sizes $k$ {(also called $k$mers as a rod occupies 
$k$ consecutive sites)} on periodic $L \times L$ square lattices. 
In simulation, the system is treated as a random sequential adsorption.
By comparing critical values of $R_2$ with the theoretical curve of $R_2$ for standard percolation 
on rectangular-shaped square lattices ($t=0$)~\cite{diFrancesco87,Pinson94,Ziff99,supMat},
we extract the dependence of $\rhoe$ on $k$, as plotted in Fig.~\ref{fig:kmers-isoradial}(a),
which suggests $\rhoe \simeq 0.4 k$ for large $k$.
The critical values of $Q_1$ and $Q_s$ for the aligned rigid rods
are also found to match those for site percolation 
on square lattices of size $(\rhoe L) \times L$, as plotted in Fig.~S11(b) of the Supplemental Material~\cite{supMat}.
These explain the continuously varying dimensionless quantities found for aligned rigid rods in Refs.~\cite{Tarasevich12,Longone2012,Longone19}.

We then consider bond percolation with anisotropic bond-placing rules.
For the triangular lattice [Fig.~\ref{fig:kmers-isoradial}(b)], three edges of a triangle 
are occupied with different probabilities $p_0$, $p_1$, and $p_2$,
and the critical probabilities satisfy $p_0+p_1+p_2-p_0p_1p_2=1$~\cite{Sykes1963,SykesEssam1964}.
Recently the method of isoradial graphs~\cite{Kenyon2004,GM2014,Copin2018} was developed to 
prove the equivalence of critical exponents between anisotropic and isotropic systems,  
but it has not been applied to {give values of} dimensionless quantities.
We expect that, with a shear
transformation described by $\rhoe$ and an effective boundary twist $t_{\rm e}$,
a critical dimensionless quantity takes the same value on 
isoradial graphs. 
After the isoradial mapping, the length of each edge is adjusted 
to compensate for its weight to make the system conformally invariant 
in the scaling limit~\cite{Copin2018}, 
and the angles are determined by the Kenyon-Grimmett-Manolescu formula~\cite{Kenyon2004,GM2014} 
as $\omega_i=3\arctan[\sqrt{3}(1-p_i)/(1+p_i)]$, $i=0,1,2$, 
as illustrated in {Fig.~\ref{fig:kmers-isoradial}(b)}
where the rhombus on the left-hand side transforms to 
a parallelogram with $\alpha=\omega_2/2$.
Thus, {for} the triangular lattice we derive that 
the isoradial graph has
$\rho_e = \sin(\omega_2/2) \sin(\omega_1/2) / \sin(\omega_0/2)$
and $t_{\rm e} = \cos(\omega_2/2) \sin(\omega_1/2) / \sin(\omega_0/2)$.
For the square lattice, the isoradial mapping at criticality 
is sketched in Fig.~\ref{fig:kmers-isoradial}(c),
and we derive $t_{\rm e}=0$ and $\rhoe=\tan(\omega/2)$, 
with $\omega=3\arctan[\sqrt{3}(1-p)/(1+p)]$.
We have numerically verified that wrapping probabilities of anisotropic bond percolation
on the triangular and square lattices are equal to theoretical values of standard percolation 
with the above $\rhoe$ and $t_{\rm e}$~\cite{supMat}.

It was shown recently that an effective shear transformation 
also relates the anisotropic Ising model on a square to 
the isotropic Ising model on a parallelogram~\cite{Dohm2019,Dohm2021,Dohm2021b}.
In the $q$-state Potts model~\cite{Wu1982,Arguin2002}, 
which can be represented as correlated bond percolation
by the Kasteleyn-Fortuin transformation, 
bond percolation and the Ising model 
correspond to the special cases with $q \to 1$ and $q=2$, respectively. 
We also employ the isoradial mapping to derive an effective 
shear transformation~\cite{supMat} to relate 
the anisotropic Potts model to the isotropic Potts model for 
any real value $1\leq q \leq 4$.
The validity is supported by the consistent theoretical and numerical results 
for wrapping probability $R_2$ in Table S4 of the Supplemental Material~\cite{supMat}.
This considerably extends the results for the Ising model 
which were obtained by a method combining anisotropic $\phi^4$ theory and exact correlation functions 
of the Ising model~\cite{Dohm2019,Dohm2021,Dohm2021b}.

In short, we have shown that a nontrivial transformation 
relates anisotropic to isotropic percolation and Potts systems.
The anisotropy can arise from the emergent phase, 
from the shape of constituent particles, or from anisotropic bond-placing rules. 
Our study is related to but different from previous investigations 
about the dependence of critical dimensionless quantities 
on the shape or boundary conditions of the overall system~\cite{Langlands1992,Pinson94,Ziff99,Malakis2014}.
We anticipate that the transformation is valid for other boundary conditions, 
in higher dimensions and continuum space~\cite{Finner2019a}, 
and for anisotropy induced by other mechanisms~\cite{Otten2012},
as long as correlations are weakly anisotropic~\cite{Dohm2021}.
The method of numerically determining $\rhoe$ from a known dimensionless quantity or exactly deriving shear parameters
from isoradial graphs could be very useful for future research.
Our work can have practical applications in other equilibrium and nonequilibrium phase transitions.
For example, for the nematic transition of rods, variations of the Binder parameter 
were found but not well understood~\cite{Lopez2010,Almarza2011}.
For strongly anisotropic systems~\cite{Henkel2002,Kyriakopoulos2019} 
with distinct correlation-length exponents, $\nu_{\parallel}$ and $\nu_{\perp}$, 
in the parallel and perpendicular directions,
by carefully adopting a size-dependent rescaling factor $\propto L^{\nu_{\parallel} / \nu_{\perp} - 1}$,
the dimensionless quantities can be used as a powerful tool 
in locating phase transitions~\cite{Binder1990,Winter2010,Angst2012,Norrebrock2019}.

\begin{acknowledgments}
	This work was supported by the National Natural Science Foundation of China 
	under Grants No.~11905001 (H.H.) and No.~12275263 (Y.D.), 
	and by the Anhui Provincial Natural Science Foundation under Grant No.~1908085QA23 (H.H.). 
	We acknowledge the High-Performance Computing Platform of Anhui University for providing computing resources. 
\end{acknowledgments}

\bibliographystyle{apsrev4-1}
\bibliography{Janus_revised.bbl}

\end{document}


\title{Supplemental Material for ``Universal Critical Behavior of Percolation in Orientationally Ordered Janus Particles
and Other Anisotropic Systems"}
\author{Hao Hu}
\email{huhao@ahu.edu.cn}
\affiliation{School of Physics and Optoelectronic Engineering, Anhui University, Hefei 230601, China}
\author{Robert M. Ziff}
\affiliation{Center for the Study of Complex Systems and Department of Chemical Engineering, University of Michigan, Ann Arbor, Michigan 48109-2800, USA}
\author{Youjin Deng}
\affiliation{Department of Modern Physics, University of Science and Technology of China, Hefei 230026, China}
\affiliation{MinJiang Collaborative Center for Theoretical Physics, College of Physics and Electronic Information Engineering, Minjiang University, Fuzhou 350108, China}
\maketitle

\newcommand{\beginsupplement}{%
        \setcounter{table}{0}
        \renewcommand{\thetable}{S\arabic{table}}%
        \setcounter{figure}{0}
        \renewcommand{\thefigure}{S\arabic{figure}}%
        \setcounter{equation}{0}
        \renewcommand{\theequation}{S\arabic{equation}}%
     }
\onecolumngrid
\beginsupplement

In this supplemental material, Sec.~\ref{sec.Janus} contains details for the system of Janus disks on the triangular lattice, 
including details of the thermodynamic phase transition and the percolation transition. Section~\ref{sec.rods} presents
details for the system of aligned rigid rods. Section~\ref{sec.bond} gives detailed results for anisotropic bond percolation, which verify the theoretical results
obtained using the method of isoradial graphs. The shear parameters for the honeycomb lattice are also given by using the star-triangle transformation. Section~\ref{sec.wrapping} presents a script in Mathematica for calculating 
exact values of wrapping probabilities for standard percolation in two dimensions using expressions from the literature.
Finally, Section~\ref{sec.aniPotts} includes our preliminary results for anisotropic $q$-state Potts model on the triangular lattice.

 \section{For Janus disks on the triangular lattice}
 \label{sec.Janus}

 \subsection{Details of the thermodynamic phase transition of the Janus system}

We performed Monte Carlo (MC) simulations at different points $(\theta,T,L)$ for interacting Janus disks of half-patch angle $\theta$ at temperature $T$ on $L \times L$ rhombus-shaped triangular lattices
with periodic boundary conditions. The Metropolis algorithm was employed, for which {the disks are visited sequentially, and} at each step a Janus disk is proposed to rotate randomly in $[-\pi,\pi)$.
For each run of the simulations, $1/3$ of the whole Markov chain was employed for thermalization, and the remaining $2/3$ was used for sampling. 
The typical length of a chain in a single run was around $3\times10^7$ sweeps, with one sweep consisting of $L^2$ single steps.
We checked that the thermalization time is much longer that than the autocorrelation time.
Dozens of runs were performed simultaneously for a same set of $(\theta,T,L)$ to gain adequate samples in a reasonable time.
We simulated sizes up to $L=128$ for most points $(\theta,T)$, and added simulations at $L=256$ for a few points.
We sampled same equilibrium quantities, e.g., the heat capacity, stripe order parameter and associated Binder parameter,  
as defined in Ref.~\cite{Mitsumoto18}.
{
The order parameter is $|\bm{\psi}|=\sqrt{\psi_x^2 + \psi_y^2}$, 
with 
\begin{equation}
	\psi_x=\frac{\sqrt{3}}{2} \left(S(\bm{k_2})-S(\bm{k_3})\right) \,, \\
	\psi_y=S(\bm{k_1}) - \frac{1}{2}\left(S(\bm{k_2})+S(\bm{k_3})\right) \,,
	\nonumber
\end{equation}
where $S(\bm{k}_i)$ is the structure factor, $\bm{k}_1$, $\bm{k}_2$ and $\bm{k}_3$ indicate three possible orientations
of the stripes~\cite{Mitsumoto18}. The Binder parameter is given by 
\begin{equation}
	B=\frac{3}{2}\left(1-\frac{\langle \psi^4 \rangle}{3 \langle \psi^2 \rangle^2 }\right)\,.
	\nonumber
\end{equation}
}

\begin{figure}[htbp]
\centering
\includegraphics[width=3.2in]{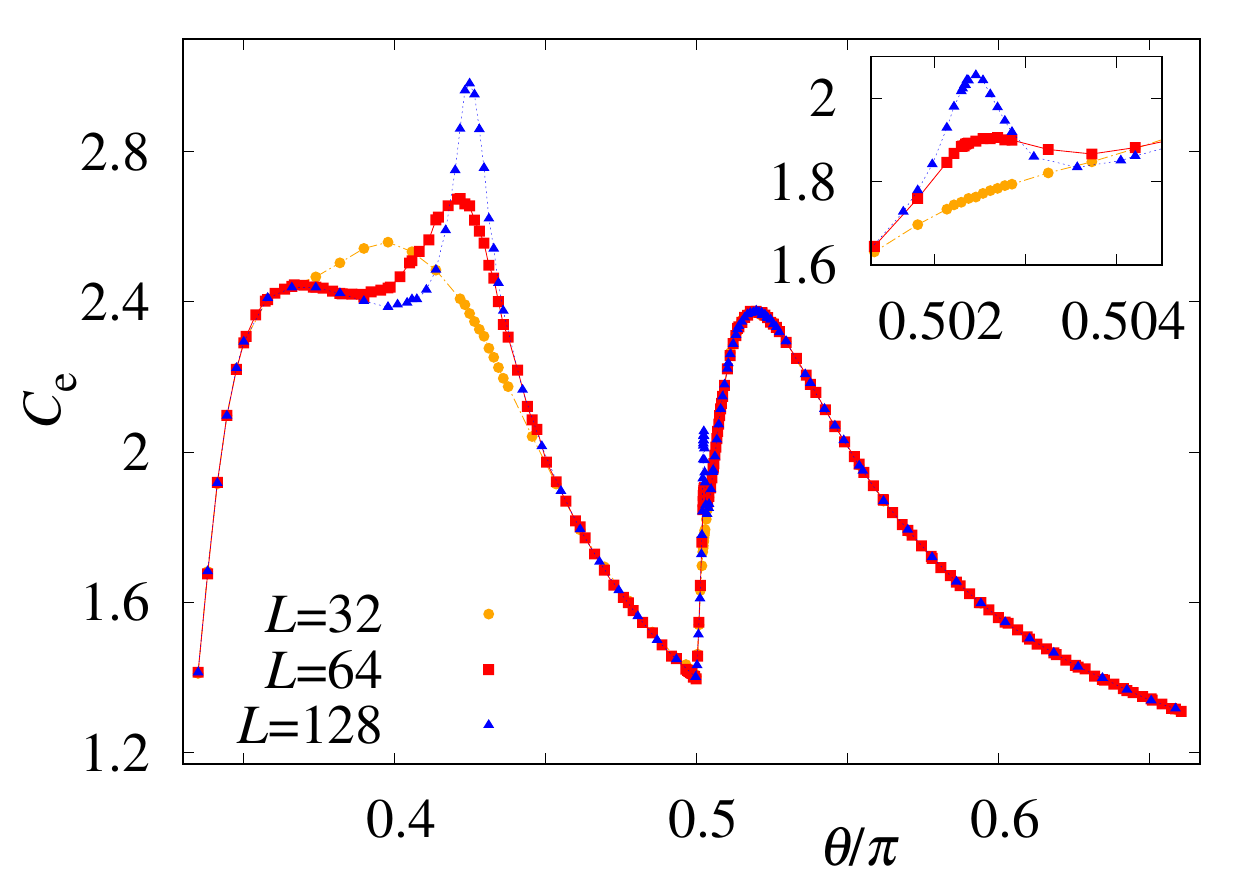}
        \caption{The heat-capacity $C_{\rm e}$ of the Janus system vs $\theta/\pi$ at $T=0.25$,
        for $\theta/\pi \in (1/3, 2/3)$.
        The inset is an enlargement of the main plot near $\theta/\pi=0.503$.
        Error bars are smaller than the data points,
        and lines connecting data points are added for clarity.~\label{fig.Ce}}
\end{figure}

Below we illustrate the thermodynamic phase behavior by presenting detailed results at $T=0.25$.
Figure~\ref{fig.Ce} shows the heat capacity $\Ce$ vs $\theta/\pi$ for three different system sizes $L$ at $T=0.25$.
At each $L$, the curve shows four different peaks. The two broad peaks near $\theta/\pi=0.36$ and $0.52$
have no appreciable dependence on the system size, thus they should indicate crossover behavior,
instead of thermodynamic phase transitions. In contrast, the other two peaks near $\theta=0.43$ and $0.503$
have significant finite-size dependence and tend to diverge as the system size increases, suggesting
thermodynamic phase transitions. The phase transition near $\theta=0.43$ is rightly the one studied 
by Mitsumoto and Yoshino~\cite{Mitsumoto18}, and the phase transition for $\theta/\pi>1/2$ has not
been reported before.

\begin{figure}[htbp]
\centering
\includegraphics[width=4.0in]{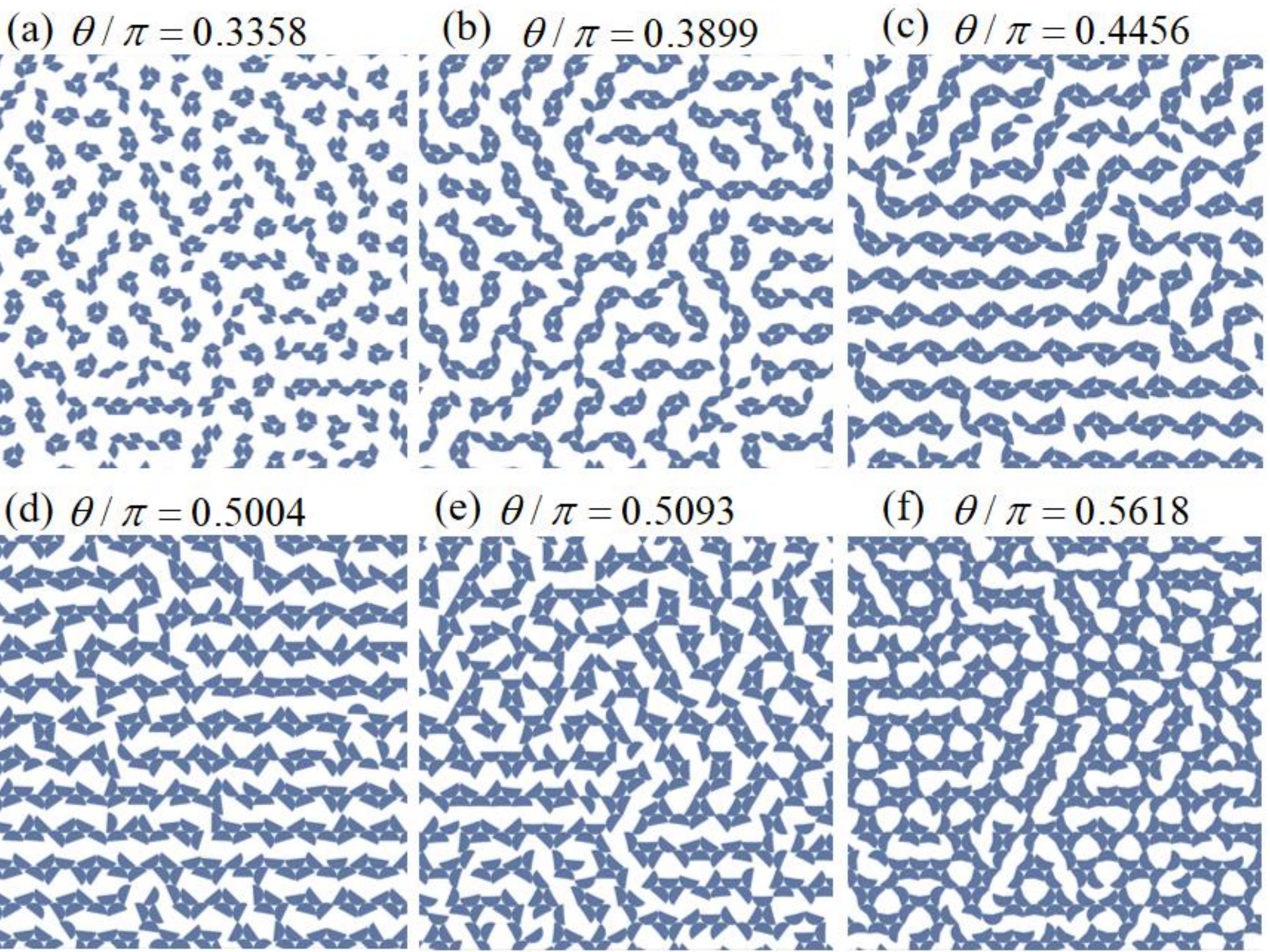}
        \caption{Snapshots for equilibrium configurations of the Janus system at $L=64$ and $T=0.25$.~\label{fig.snapshots}}
\end{figure}

\begin{figure}[htbp]
\centering
\includegraphics[width=3.2in]{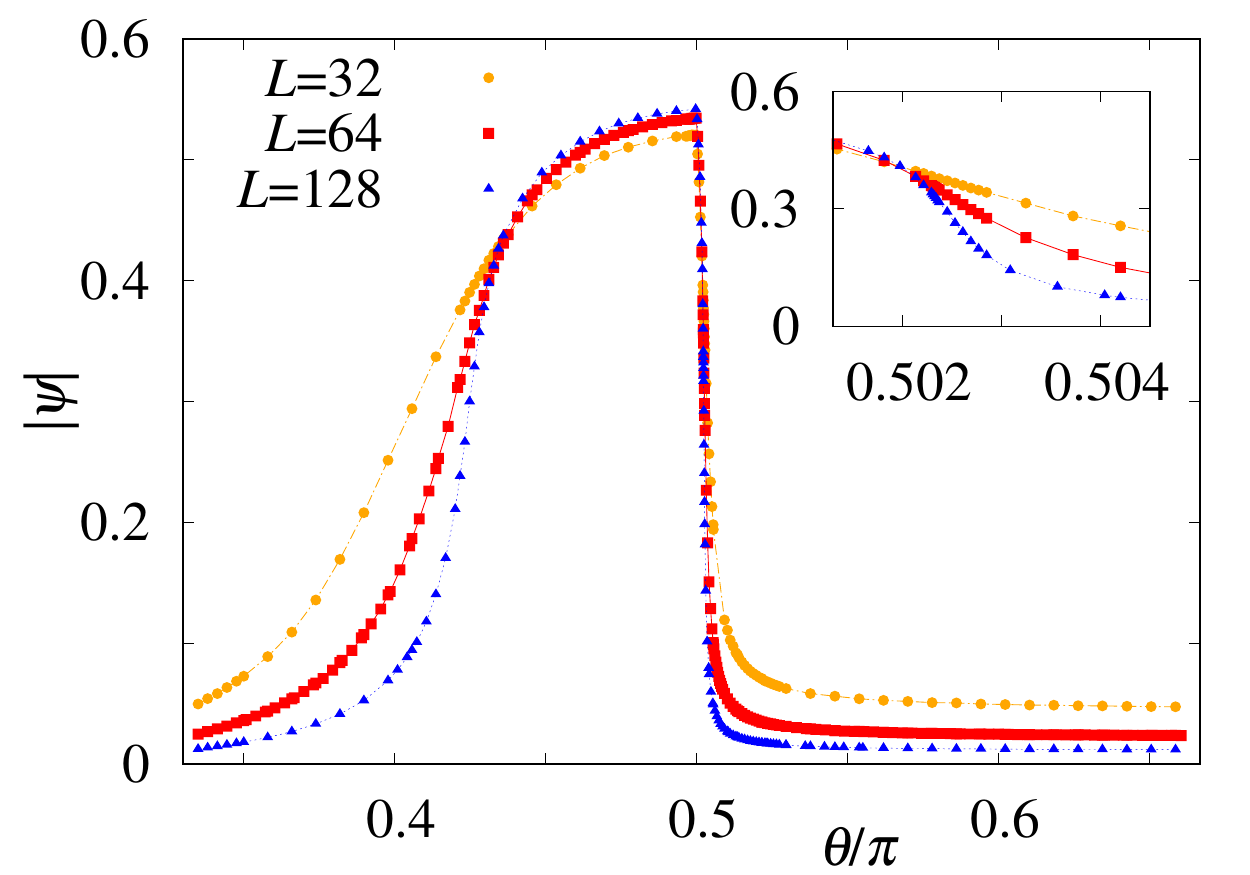}
        \caption{The stripe order parameter $|\psi|$ vs $\theta/\pi$ at $T=0.25$,
        for $\theta/\pi \in (1/3, 2/3)$.
        The inset is an enlargement of the main plot near $\theta=0.503$.
        Error bars are smaller than the data points,
        and lines connecting data points are added for clarity.~\label{fig.Ps}}
\end{figure}

We plot configurations for $T=0.25$ at different $\theta/\pi$ values,
as shown in Fig.~\ref{fig.snapshots}. One can regard the above peak near $\theta/\pi=0.36$ as the effect
of crossover between a region with many trimers [e.g., Fig.~\ref{fig.snapshots}(a)]
and a region with large polymers [e.g., Fig.~\ref{fig.snapshots}(b)].
The peak near $\theta/\pi=0.52$ can be considered as the effect of crossover between
a region where the white space consists of many white trimers [e.g., Fig.~\ref{fig.snapshots}(f)]
and a region where the white space has many larger white polymers [e.g., Fig.~\ref{fig.snapshots}(e)].
The peak near $\theta/\pi=0.43$ signifies a phase transition from
a disordered phase [e.g., Fig.~\ref{fig.snapshots}(b)] into a stripe phase [e.g., Fig.~\ref{fig.snapshots}(c)]
which breaks a global three-fold rotation symmetry~\cite{Mitsumoto18}.
We find that the ordered stripe phase also exists for $\theta>\pi/2$ [e.g., Fig.~\ref{fig.snapshots}(d)],
and through a phase transition near $\theta/\pi=0.502\,3$, the global three-fold symmetry is
recovered [e.g., Fig.~\ref{fig.snapshots}(e)]. The stripe order between the two peaks near $\theta/\pi=0.43$ and
$\theta/\pi=0.503$ is confirmed by the plot of the order parameter $|\psi|$~\cite{Mitsumoto18} in Fig.~\ref{fig.Ps}.

\begin{figure}[htbp]
\centering
\includegraphics[width=3.2in]{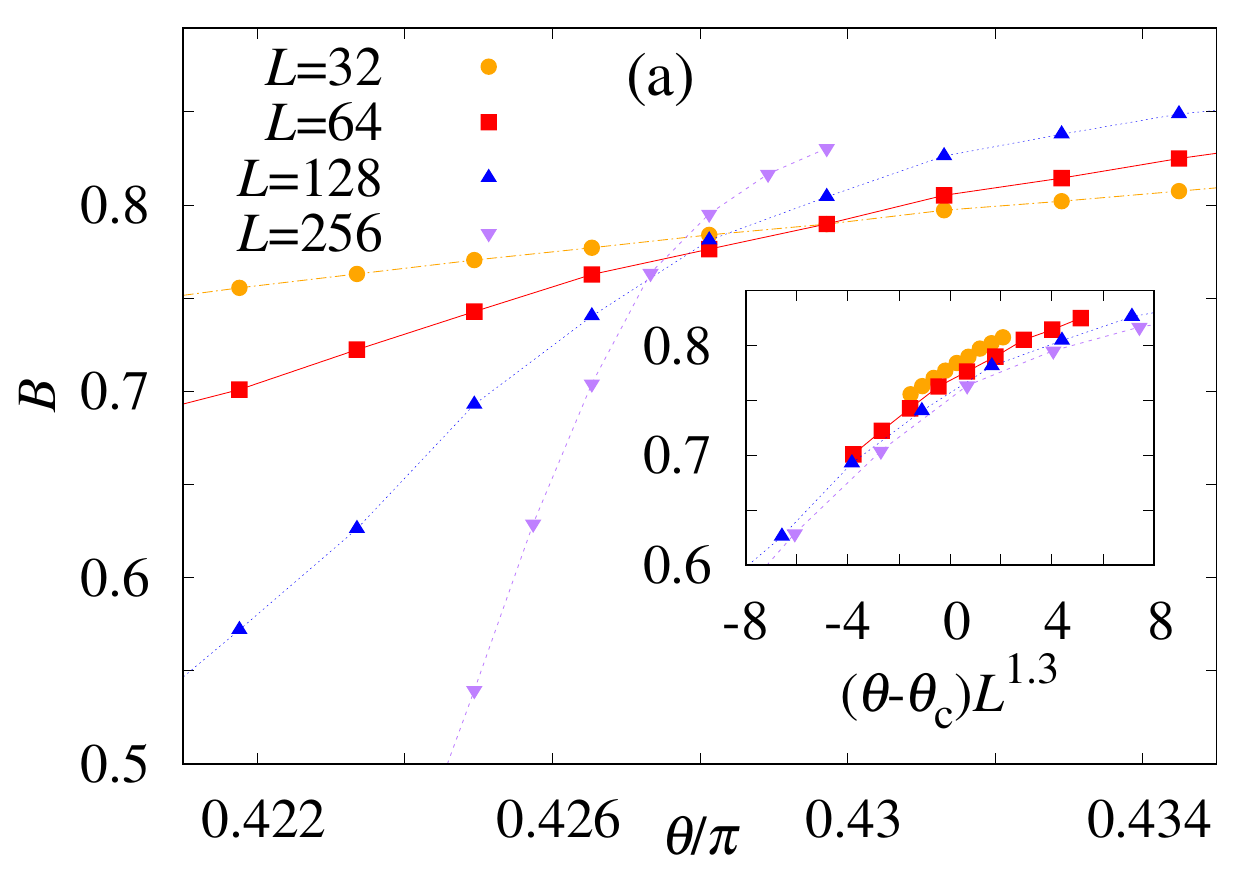} 
\includegraphics[width=3.2in]{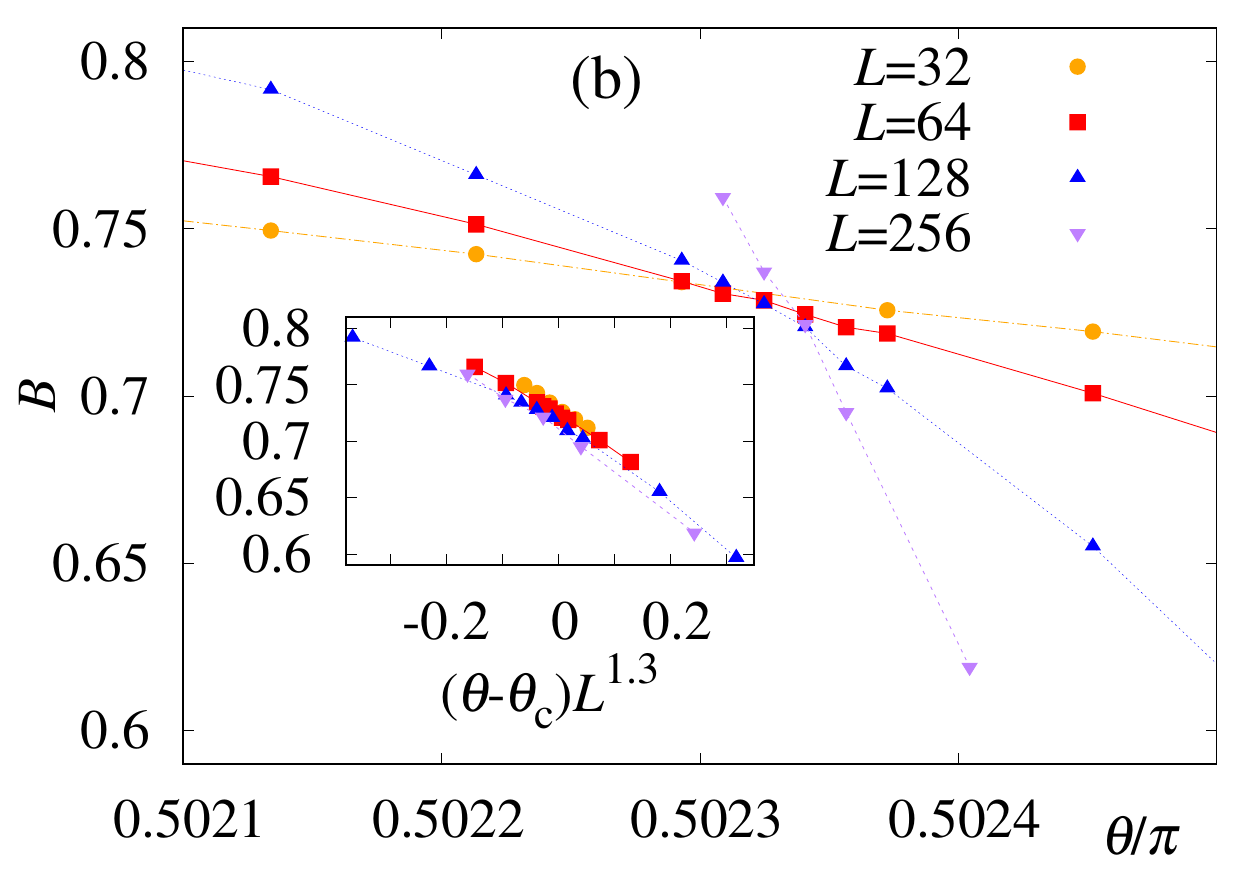}
    \caption{The Binder parameter $B$ vs $\theta/\pi$ near the two phase transition points $\theta_{\rm c, 1}/\pi \simeq 0.427$ (a) and $\theta_{\rm c, 2}/\pi \simeq 0.502\,4$ (b) at $T=0.25$ for the Janus system. The insets show data collapse for the main plots, i.e., $B$ vs $(\theta-\theta_{\rm c})L^{1.3}$, where the exponent $1.3$ comes from the fits.~\label{fig.Bi}}
\end{figure}

The above transition near $\theta/\pi=0.43$ was suggested to be a continuous phase transition~\cite{Mitsumoto18}. 
In Ref.~\cite{Mitsumoto18}, in addition to the continuous change of the order parameter, 
the continuity of the phase transition for $\theta/\pi<1/2$ is further supported 
by a crossing feature of the Binder  $B$, as exemplified in Fig.~\ref{fig.Bi}(a).
For the phase transition near $\theta/\pi=0.503 > 1/2$, in addition to the continuous change of 
the order parameter as shown in Fig.~\ref{fig.Ps}, we find that the data of $B$
at different system sizes also cross near the transition point. These results strongly suggest that this transition near $\theta/\pi=0.503>1/2$ 
is also continuous. From the crossing of $B$, we estimate the two transition points
to be $\theta_{\rm c, 1}/\pi \simeq 0.427$ and $\theta_{\rm c, 2} / \pi \simeq 0.502\,4$.
Using the Binder parameter $B$, the phase transition point at other temperatures were also determined,
as plotted in the phase diagram in Fig.~1 of the main text.
Since the stripe phase is still the ground state up to $\theta/\pi=2/3$~\cite{Shin2014},
the thermodynamic phase transition line is expected to approach $\theta/\pi=2/3$ in the low-temperature limit.
It is noted that one can also fix the patch size $\theta$ and change the temperature $T$ to
determine a phase transition point.

When $\theta/\pi<1/2$, in Ref.~\cite{Mitsumoto18}, near the continuous phase transition point, approximate collapse of the data for the order parameter, its second moment and the Binder parameter was observed using critical exponents $\nu=5/6$, $\beta=1/9$ and $\gamma=13/9$ of 
the universality class of $3$-state Potts model~\cite{Wu1982}. 
For $\theta/\pi>1/2$, from the disordered phase to the ordered stripe phase, since a global three-fold rotation symmetry
is also broken as for $\theta/\pi<1/2$~\cite{Mitsumoto18}, it is expected that the phase transition is {in} the same universality class.
We fitted the data of $B$ at $T=0.25$ near the crossing points to the finite-size scaling ansatz
\begin{equation}
	B = B_0 + a_1(\theta-\theta_c)L^{y_{\rm t}} + a_2(\theta-\theta_c){^{2}}L^{2y_{\rm t}}+b_1L^{y_{\rm i}} \,,
        \label{eq:Q}
\end{equation}
where $B_0$ represents the value of $B$ in the thermodynamic limit, $y_{\rm t}=1/\nu$ is the thermal renormalization exponent, $y_{\rm i}<0$ is the leading irrelevant scaling exponent,
$a_i$ and $b_i$ are nonuniversal amplitudes.
With $y_{\rm i}$ being fixed at $-1$ or $-2$ and using data points with $L \ge 64$, from the fits we obtain effective values
of the exponent $y_{\rm t}$ as $1.29(8)$ and $1.36(13)$ for the transitions near $\theta_{\rm c} / \pi \simeq 0.427$ and $0.502\,4$,
respectively. When plotting $B$ vs $(\theta-\theta_{\rm c})L^{1.3}$, approximate data collapse can be obtained
as in the insets of Fig.~\ref{fig.Bi}.
The slight deviation of $y_{\rm t}=1/\nu$ from the $3$-state Potts value $6/5$ either comes from finite-size effects or hints another universality class.
The latter possibility remains to be explored, since in terms of symmetry, the transitions to the stripe phase break a $Z_3$ rotation symmetry,
while phase transition in the $3$-state Potts model breaks a $S_3$ permutation symmetry.

 \subsection{Details of the percolation transition of the Janus system}

 When studying the percolation transition, the configuration of the Janus system was still updated using the Metropolis algorithm.
 For each run of the simulations, the equilibrium stage took around $10^7$ sweeps, and the sampling stage took another $10^7$ (or more) sweeps. 
 In the sampling stage, measurements were performed every $10$ (or more) sweeps. For measuring a configuration,  a breadth-first search was 
 conducted to count cluster sizes and to determine whether or how the clusters wrap. 
 We simulated sizes up to $L=256$ for most points $(\theta,T)$, and added simulations at $L=512$ for a few points.
 Around ten runs were performed simultaneously for a same set of $(\theta,T,L)$ to gain adequate samples in a reasonable time.

 In the following, we first demonstrate that, for percolation in the Janus system, the critical exponents 
 are consistent with standard percolation. 
 Then we illustrate the determination of dimensionless quantities.
 We also show how we use an appropriate rescaling to collapse the critical correlations {at another temperature (besides $T=0.23$ in the main text)}.
 In the end we present the change of local connectivity along the percolation line.

\begin{figure}[htbp]
\centering
        \includegraphics[width=3.2in]{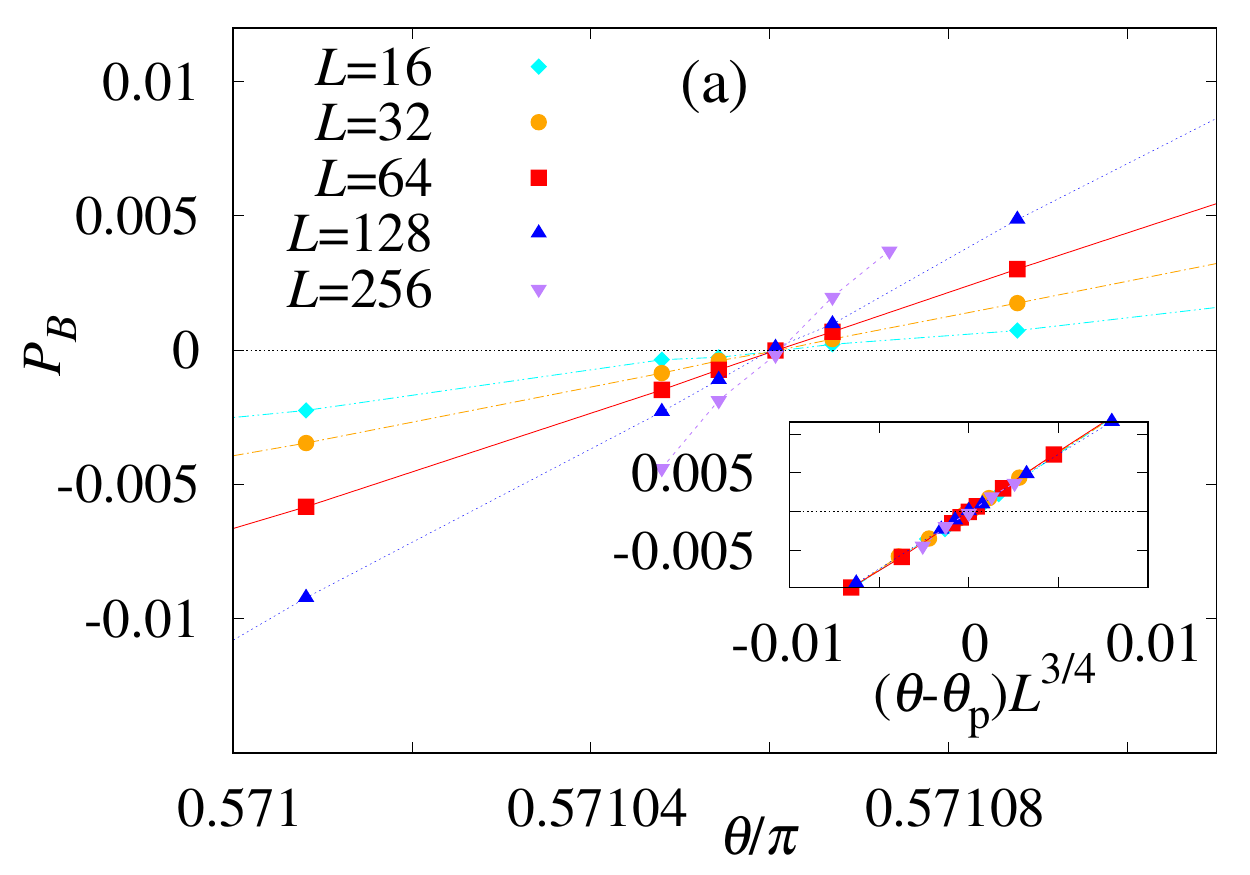}
	\includegraphics[width=3.2in]{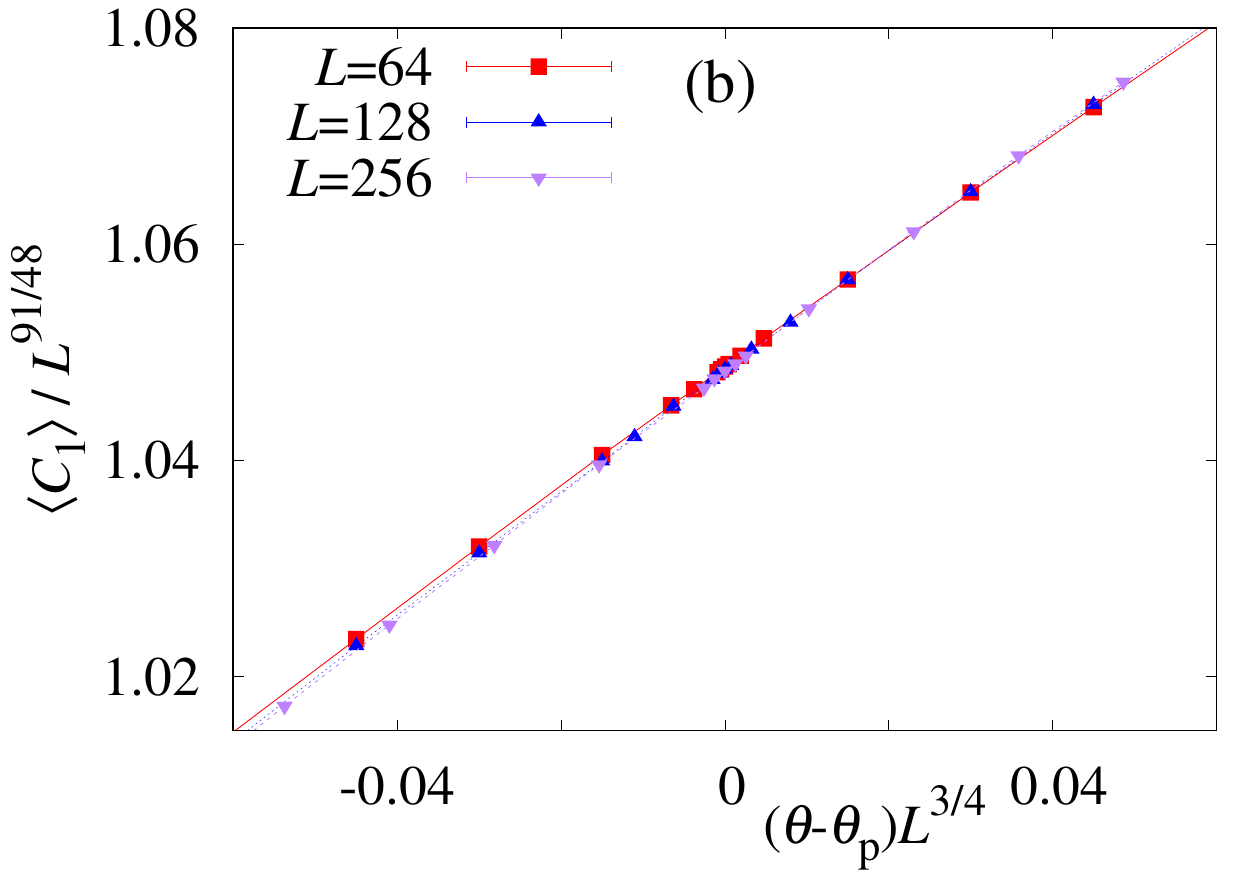}
	\caption{Results for percolation in the disordered phase of the Janus system at $T=1$ near the percolation
	threshold $\thetap/\pi=0.571\,061$: (a) Plot of $\PB$ vs $\theta/\pi$ 
	[vs $(\theta - \theta_{\rm p})L^{3/4}$ in the inset]; 
	(b) Plot of $\langle C_1 \rangle/L^{91/48}$ vs $(\theta - \theta_{\rm p})L^{3/4}$.~\label{fig.T1}}
\end{figure}

 Figure~\ref{fig.T1} shows simulation results at $T=1$ for percolation in the disordered phase of the Janus system.
 The data of $\PB$ is fitted using the least-square method by the finite-size scaling ansatz
\begin{equation}
	\PB = O_0 + a (\theta-\thetap)L^{\yt} + b L^{\yi} \,\,,
        \label{eq:PB}
\end{equation}
where $O_0$ is expected to be zero for standard percolation,
$\thetap$ is the percolation threshold, $\yt=1/\nu$ is the thermal renormalization exponent,
and $\yi<0$ is the leading irrelevant exponent.
In the fits, a lower cutoff $L_{\rm min}=16$ is taken so that higher-order corrections
can be neglected. With $O_0$ and $\yt$ being fixed or free, and $\yi$
assuming different values, the fit results are consistent with $\yt=1/\nu=3/4$ and $O_0=0$ for standard percolation,
and lead to the estimate $\thetap/\pi=0.571\,061(1)$. Well collapse of the data can be obtained as shown
in the inset of Fig.~\ref{fig.T1}(a). We also observe the finite-size scaling of the average size of the largest
cluster $\langle C_1 \rangle$, and find that it can be described by $\langle C_1 \rangle = L^{\df}(a' (\theta-\thetap)L^{\yt} + b' L^{\yi'})$
near the percolation threshold $\thetap$, with the magnetic renormalization exponent $\df = y_{\rm h}$ 
being consistent with the value $91/48$ of standard percolation.
The latter is exhibited by the plot in Fig.~\ref{fig.T1}(b).

\begin{figure}[htbp]
\centering
        \includegraphics[width=3.2in]{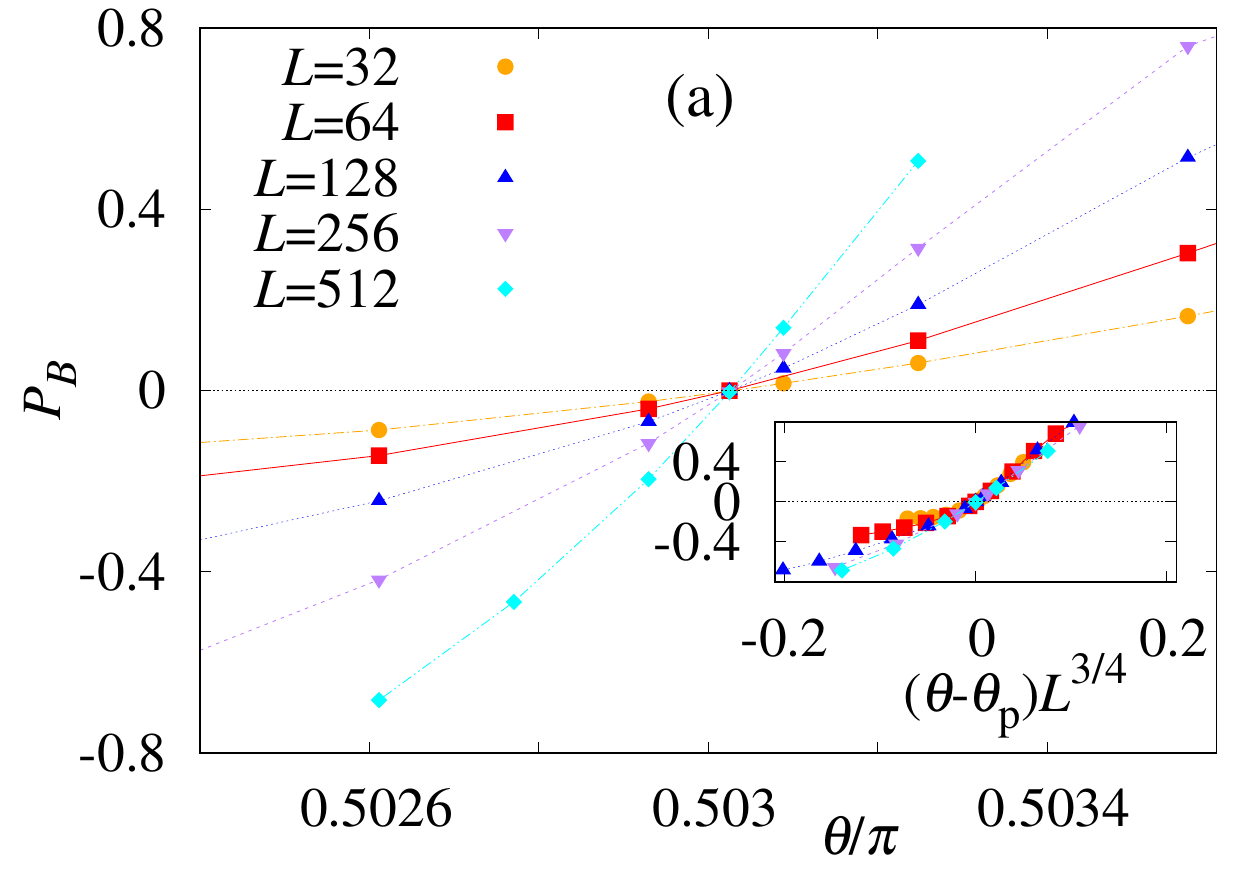}
	\includegraphics[width=3.2in]{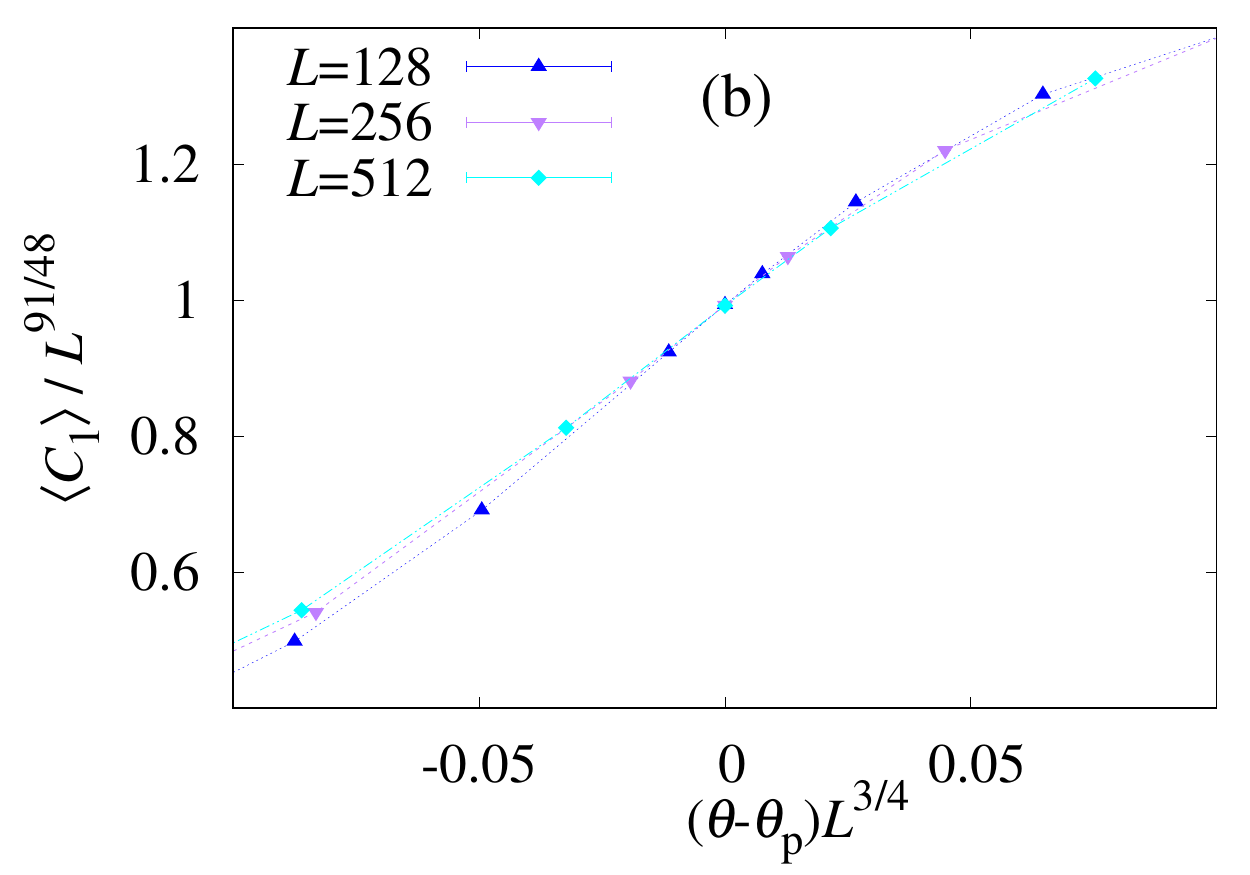}
	\caption{Results for percolation in the ordered stripe phase of the Janus system at $T=0.23$ 
	near the percolation threshold $\thetap/\pi=0.503\,025$: (a) Plot of $\PB$ vs $\theta/\pi$ 
	[vs $(\theta - \theta_{\rm p})L^{3/4}$ in the inset]; 
	(b) Plot of $\langle C_1 \rangle/L^{91/48}$ vs $(\theta - \theta_{\rm p})L^{3/4}$.~\label{fig.Tdot23}}
\end{figure}

Figure~\ref{fig.Tdot23} shows simulation results at $T=0.23$ for percolation in the ordered stripe phase of the Janus system.
The value of $\PB$ is consistent with zero at the percolation threshold,
and the critical exponents are also consistent with $y_t=1/\nu=3/4$ and $\df=y_{\rm h}=91/48$ for standard percolation.

Thus, both in the disordered phase and the ordered stripe phase, the percolation transition in the Janus system
belongs to the universality class of standard percolation. 
The critical polynomial $\PB$ for different sizes $L$ nicely cross near $\thetap$ with $\PB(\thetap,L\rightarrow\infty)=0$,
and it was used to determining the whole percolation line, as plotted in Fig.~1 of the main text. 
We have also checked critical exponents at some other points of the percolation line and found the values being consistent 
with those of standard percolation.

\begin{figure}[htbp]
\centering
	\includegraphics[width=2.5in]{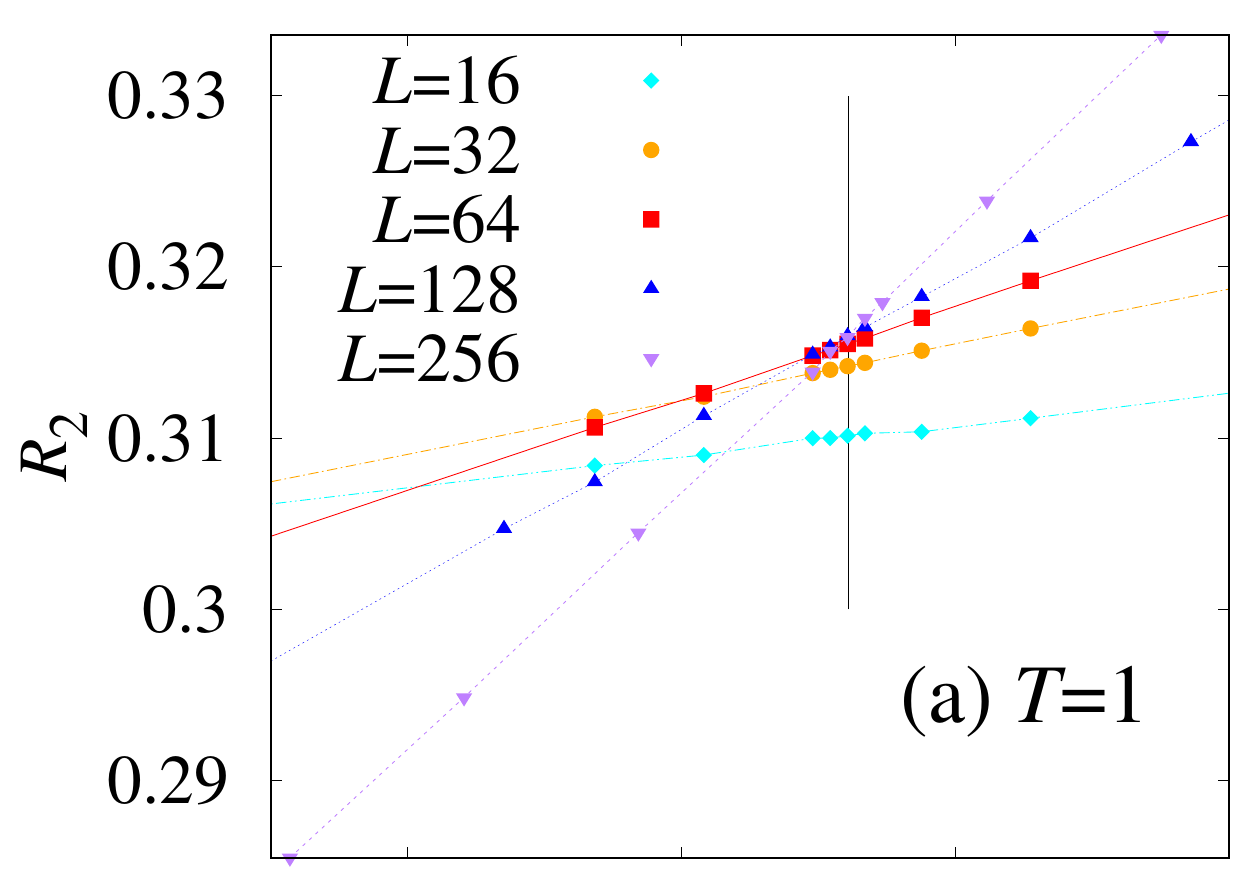} \hspace{-1.5mm}
	\includegraphics[width=2.5in]{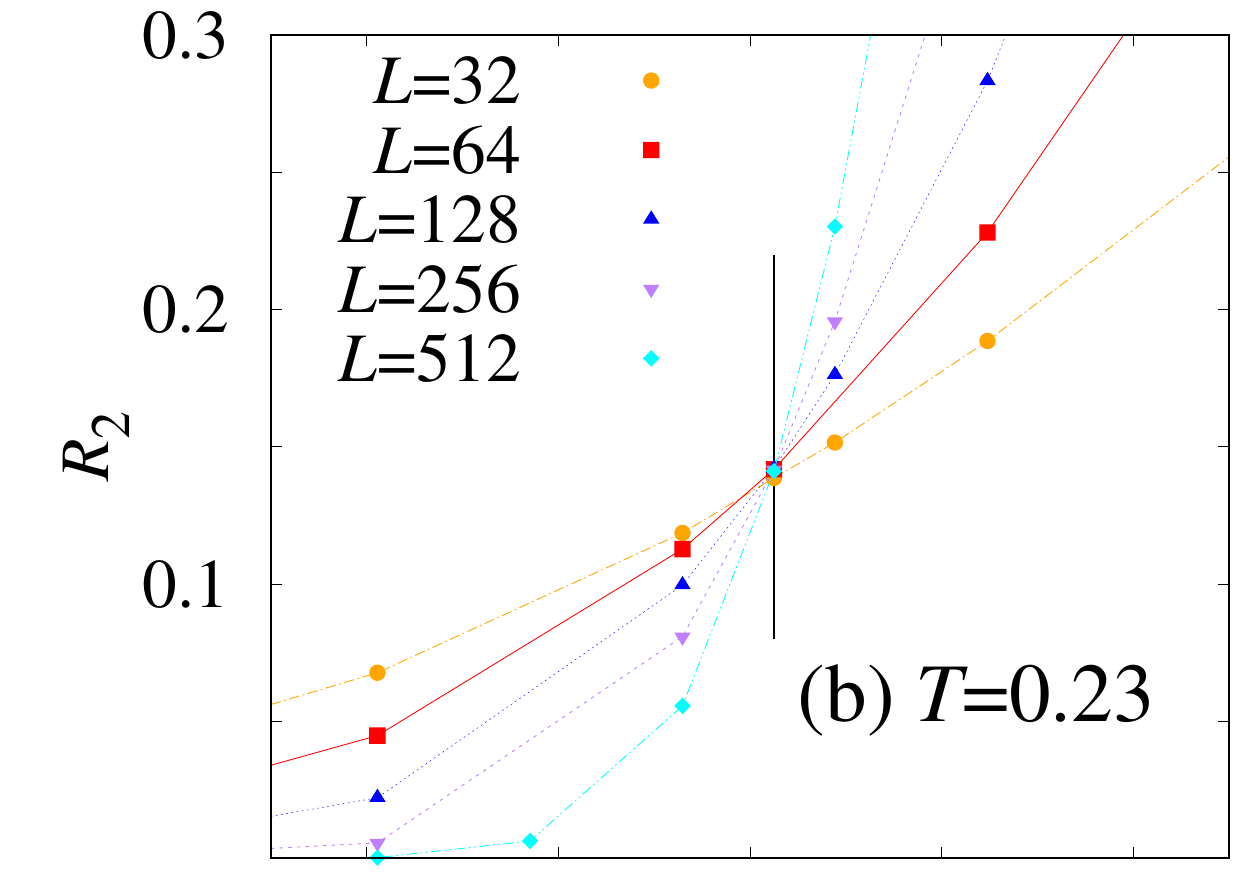} \\
	\vspace{-2.36mm}
	\includegraphics[width=2.5in]{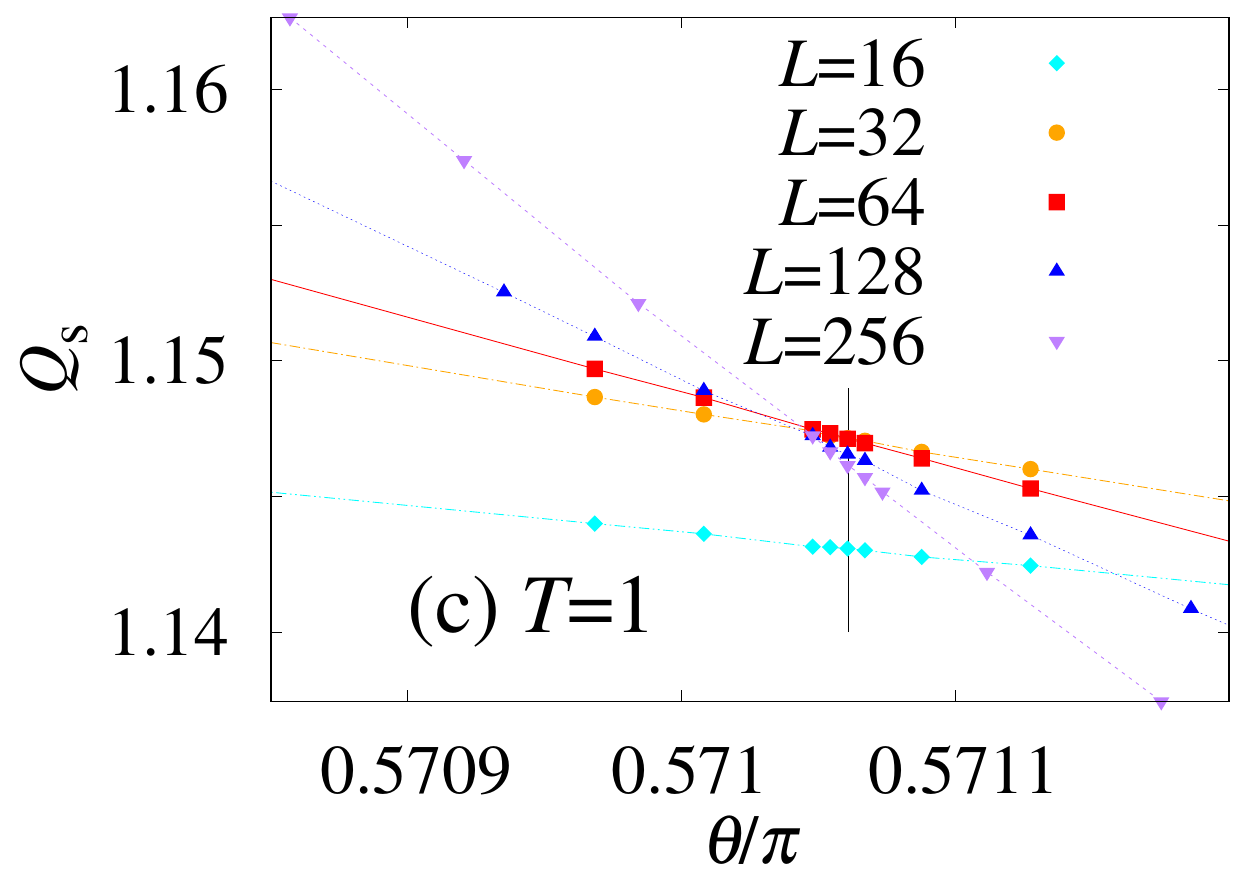} \hspace{-1.5mm}
	\includegraphics[width=2.5in]{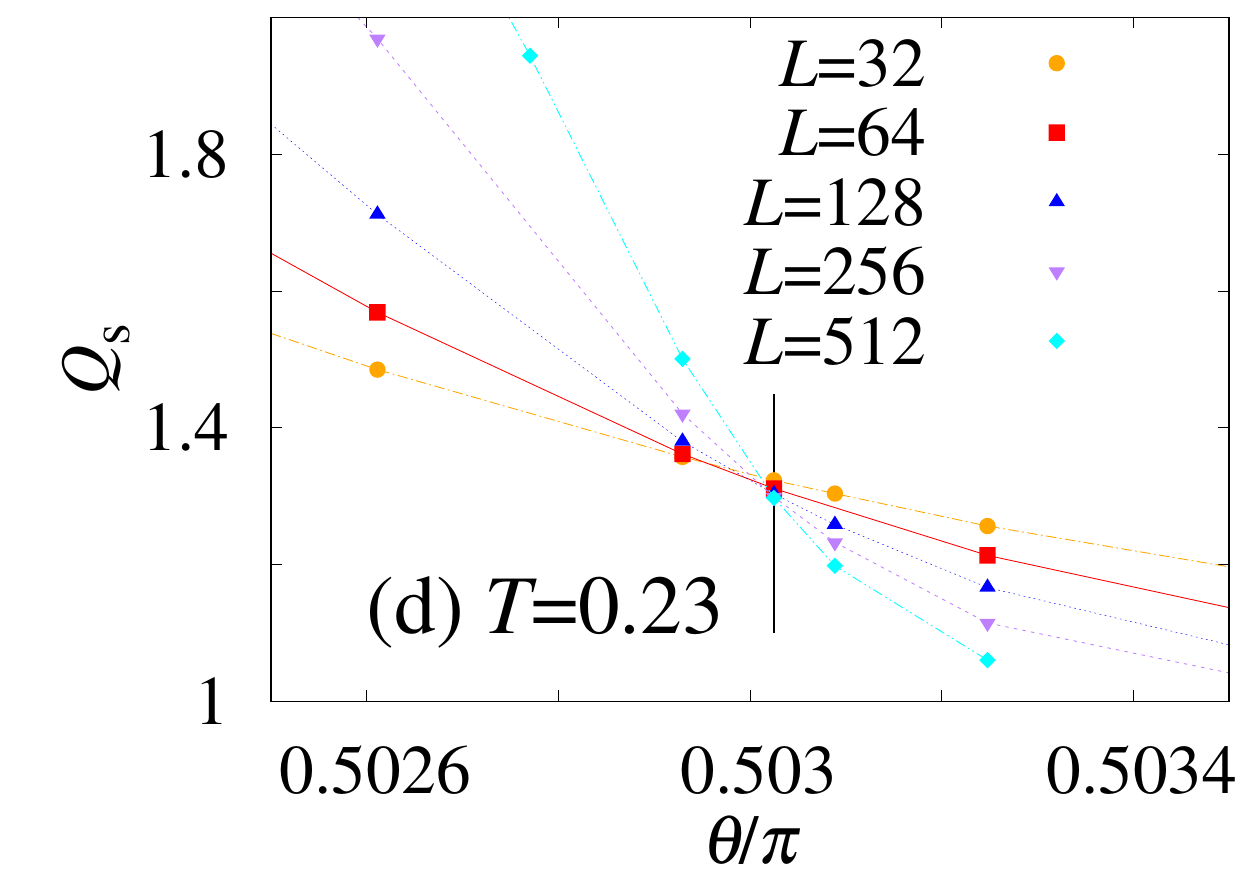} \\
	\caption{
		The wrapping probability $R_2$ and dimensionless ratio $Q_{\rm s}$ for percolation
		of the Janus system at $T=1$ (a,c) and $T=0.23$ (b,d).
		The vertical solid line indicates the position of the percolation threshold
		$\thetap/\pi=0.571\,061$ and $0.503\,25$ for $T=1$ and $T=0.23$, respectively.~\label{fig.RQ}}
\end{figure}

While the critical value of $\PB$ is consistent with zero along the whole percolation line.
As presented in the main text, values of wrapping probabilities $R$ and dimensionless ratios $Q$ change along the percolation
line in the ordered stripe phase. Figure~\ref{fig.RQ} illustrates the crossing of $R_2$ and $Q_s$ near the percolation
threshold for $T=1$ and $T=0.23$. Other $R$ and $Q$ also have similar crossing, which are not plotted for brevity.  
It can be seen that the critical values of $R$ and $Q$ for $T=0.23$ are clearly different from those for $T=1$.
At different temperatures, the precise values of $R$ and $Q$ at the percolation threshold 
are determined by fitting the data to Eq.~(\ref{eq:PB}), with $\PB$ in the equation replaced by $R$ or $Q$, and $O_0$ being nonzero.

For site percolation, as $\rho \rightarrow \infty$, from Fig.~3(d) in the main text,
one has limiting values $Q_1=1$ and $Q_s=3$.
Since $\rhoe(T=0)=\infty$, these two values also apply to the Janus system in the low-temperature limit.
In the limit $T \rightarrow 0$ for the Janus system, 
 assuming all clusters are of constant size $C_0$, it can be proved that $Q_1=1$ and $Q_s=3$: (1) $\langle C^2_1 \rangle = \langle C^2_0 \rangle = \langle C_0 \rangle^2$, thus $Q_1 = \langle C_1^2 \rangle / \langle C_1 \rangle ^2 = \langle C_0 \rangle^2 / \langle C_0 \rangle^2 = 1$. (2) $S_l = \sum_i C_i^l = n_c C_0^l$, where the number of clusters $n_c$ tends to infinity in the thermodynamic limit. Thus $Q_s = \langle 3 S_2^2 - 2 S_4 \rangle / \langle S_2 \rangle ^2 = [3 (n_c C_0^2)^2 - 2 (n_c C_0^4)]/ [(n_c C_0^2)^2] = 3 - 2/n_c = 3$, where $n_c \rightarrow \infty$ is used.

\begin{figure}[htbp]
\centering
	\includegraphics[width=3.2in]{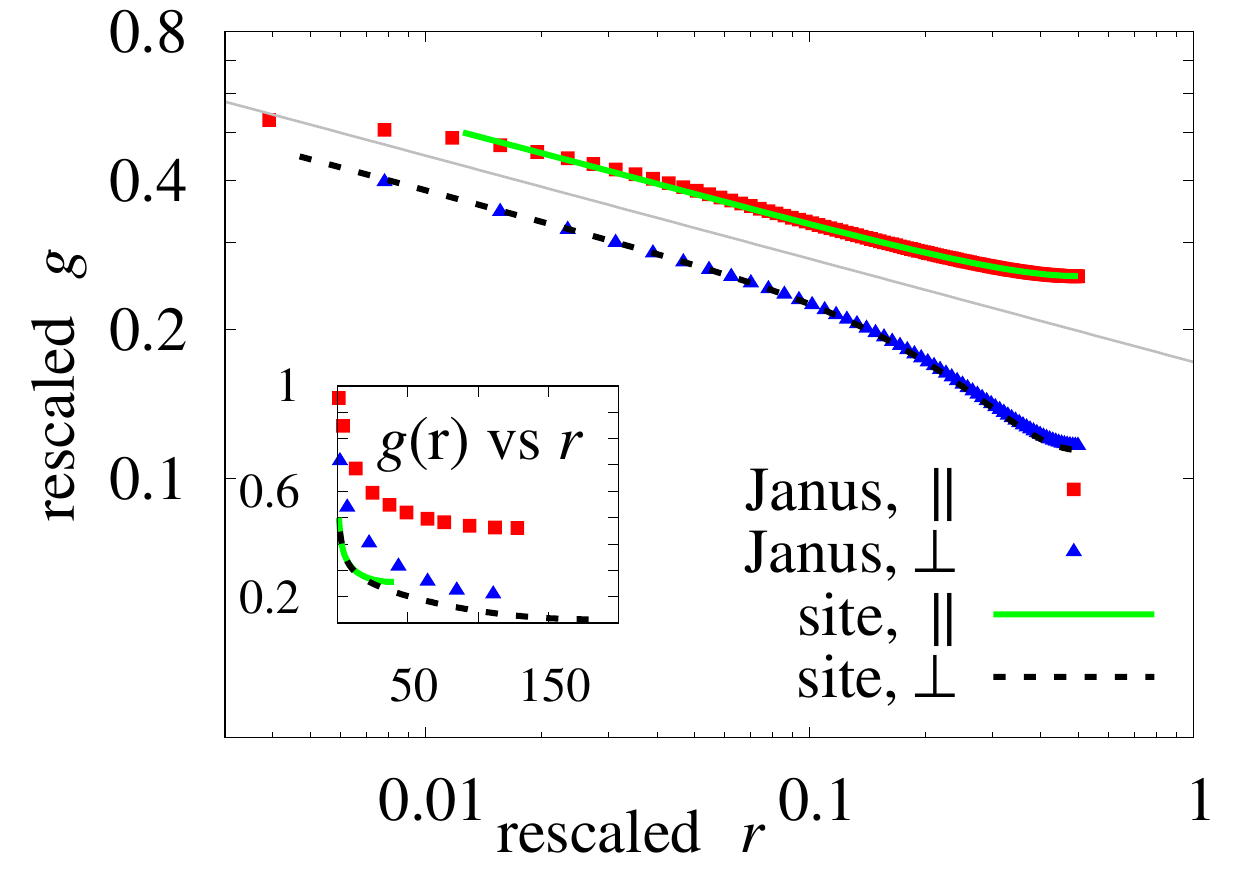}
	\caption{By {an} appropriate rescaling described in the {main text},
		 the critical correlation functions of the Janus system at $\thetap/\pi=0.502\,5$,  
		 $T=0.22$ and $L=256$ [$\rhoe=4.61(3)$]
		 in both the parallel and perpendicular directions match those of 
		 site percolation on the periodic triangular lattice with $\rho\simeq4.61$
                 (a parallelogram of size $L_{\parallel} \times L = 80 \times 426$). 
		 The slopes of the curves agree with the {standard-percolation} value $-5/24$ {(slope of the light grey line)}.~\label{fig.corrTdot22}}
\end{figure}

In the main text, {it is demonstrated that} the connectivity correlations of the Janus system and those of standard site percolation 
on the parallelogram-shaped triangular lattice are collapsed into same curves under {an appropriate rescaling.} 
{In the rescaling, a constant (being $0.588$ for $T=0.23$ in the main text) is multiplied to $\tilde{g}^{Janus}(r/L)$ to 
collapse the curves.}
Here in Fig.~\ref{fig.corrTdot22} for $T=0.22$, to collapse the curves, a constant $0.555$ is multiplied to rescale the critical correlations of the Janus system. Thus the value of the multiplying factor changes for different temperatures (i.e., it is nonuiversal), though rescaled correlation functions of the two models share same universal shapes. 

\begin{figure}[htbp]
\centering
        \includegraphics[width=3.2in]{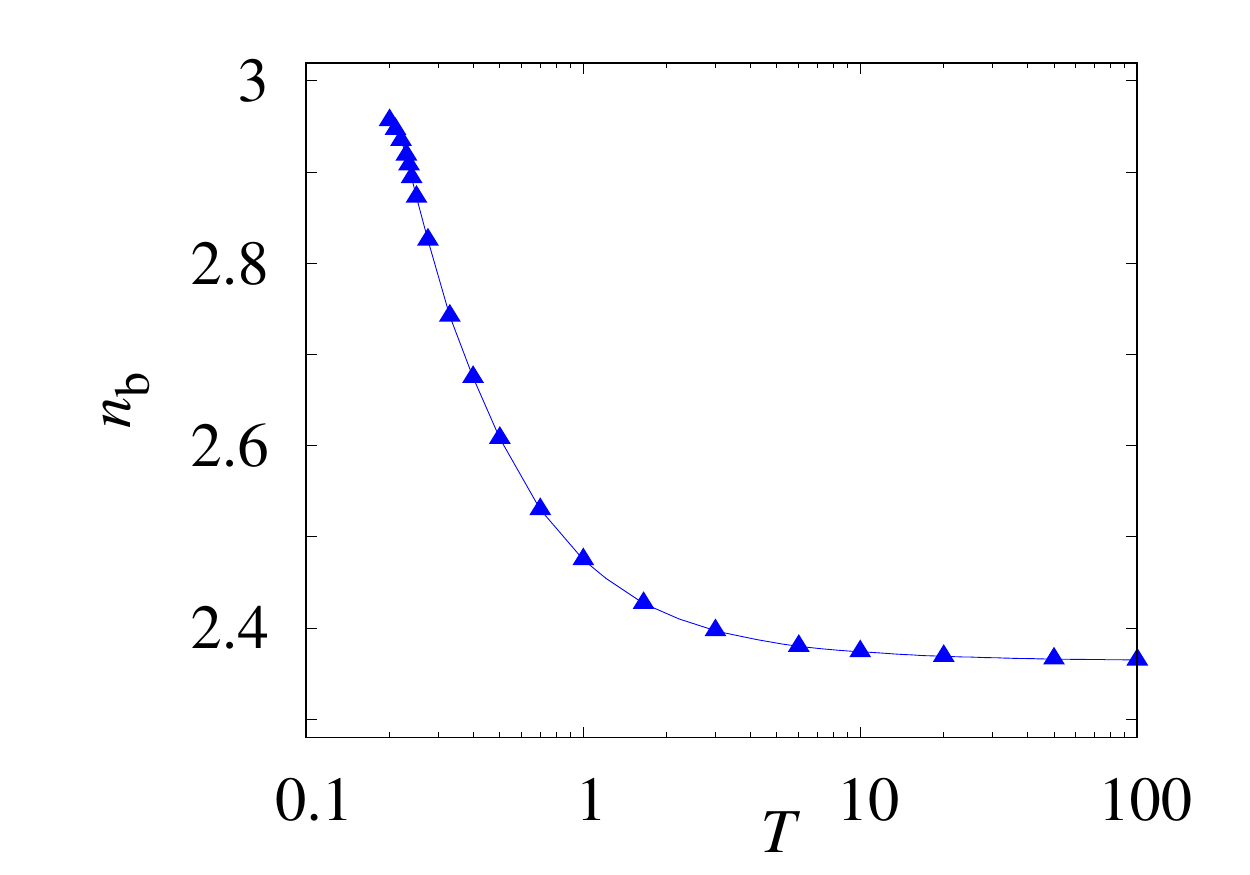}
	\caption{The average valency $n_{\rm b}$ vs $T$ along the percolation line of the Janus system.~\label{fig.valency}}
\end{figure}

The correlations and wrapping probability characterize long-range connectivity. We also sampled a quantity related to local connectivity, 
i.e., the average valency $n_b$ which is defined as the average number of connected nearest neighbors of a Janus disk.
Figure~\ref{fig.valency} shows that, as $T$ decreases, $n_b$ increases from $2.36$ at $T=100$ to near $3$  at very low temperatures.
We conjecture that, for $\theta \rightarrow \pi/2$, the percolation transition takes places at $T \rightarrow 0$,
at which $n_{\rm b} \rightarrow 3$.

 \section{For aligned rigid rods on the square lattice}
 \label{sec.rods}

 We performed MC simulations for aligned rigid rods (also called $k$-mers) on $L \times L$ periodic square lattices near the estimated percolation thresholds in Ref.~\cite{Tarasevich12}. 
 We treat the system as a random sequential adsorption in the simulation: the rigid rods are randomly put into the lattice one by one, with their ends located on lattice sites and directions oriented along the parallel ($x$) axis,
 and a rod does not overlap with any existing rod on the lattice.
 Nearest-neighboring occupied sites are treated as being connected.
 Rigid rods of sizes $k=2,4,8,12,16,20,24,28,32$ (the size is defined as the number of sites occupied by a rod) were simulated, with system sizes up to $L=100k$. 
 The number of independent samples was around $10^6$ for each $(k,L)$ at a given site occupation $p$. 
 For relating percolation of aligned rigid rods to standard percolation, simulations were also conducted for critical site percolation (corresponding to rods with $k=1$) on periodic square lattices of various aspect ratios.

\begin{figure}[htbp]
\centering
        \includegraphics[width=3.0in]{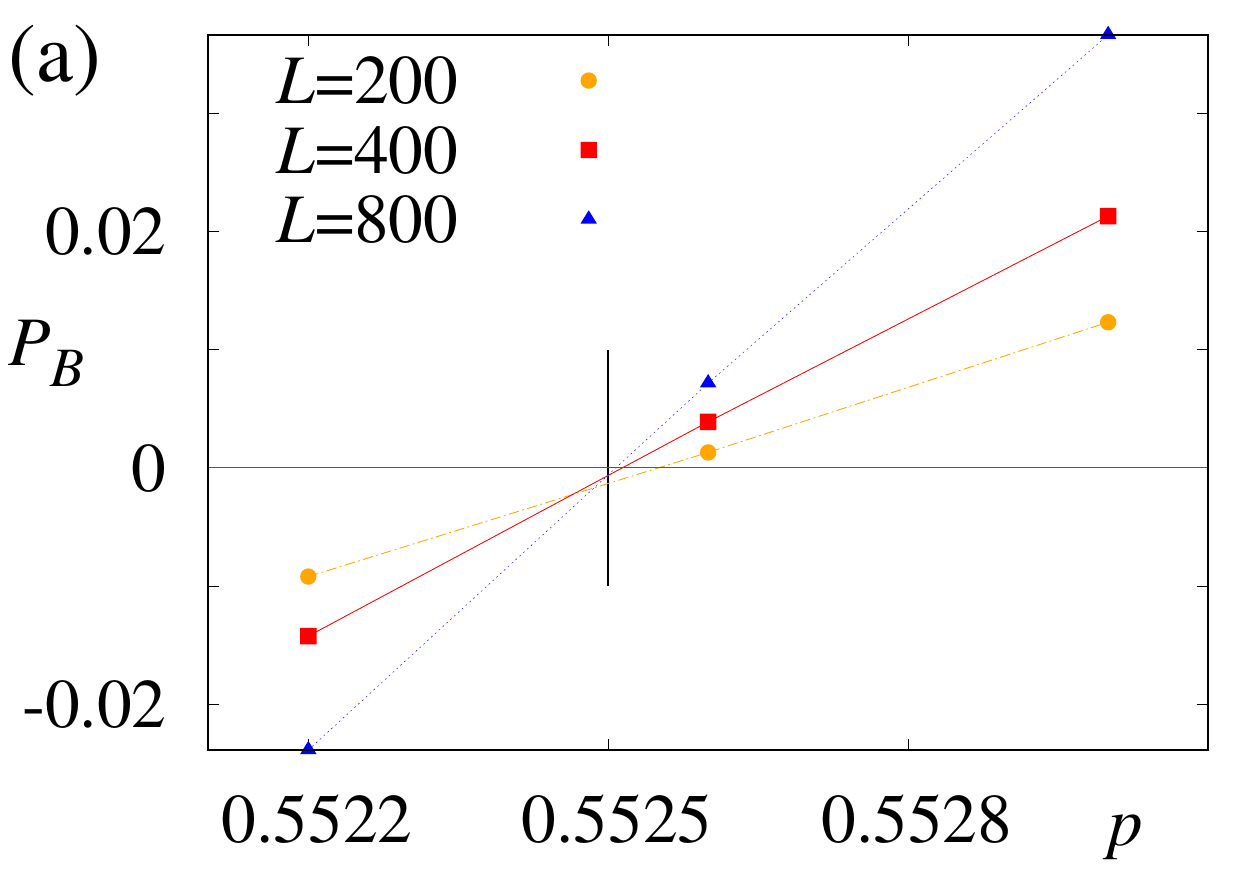}
	\includegraphics[width=3.0in]{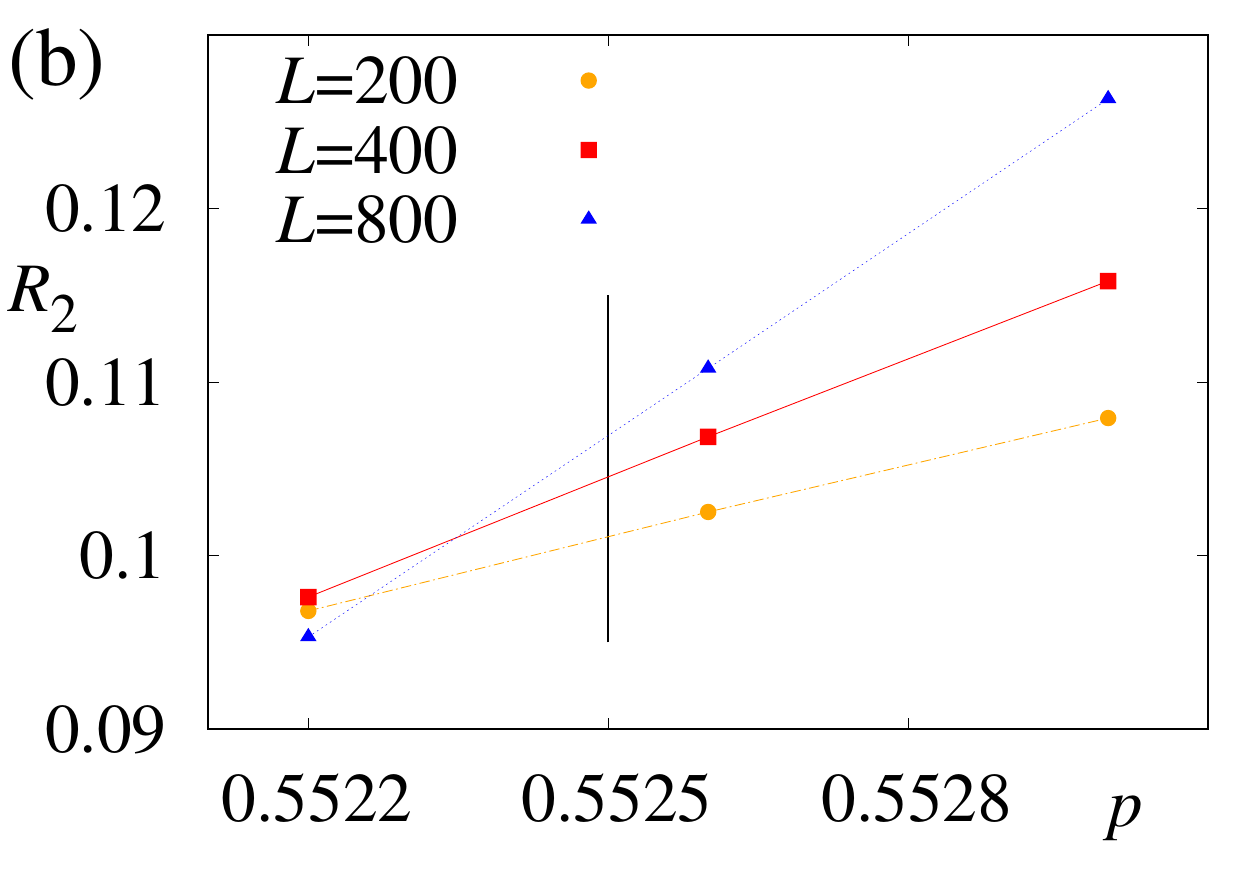}
	\caption{Results for percolation of aligned rigid rods with $k=8$ on $L \times L$ periodic square lattices:
	(a) The critical polynomial $\PB$ vs the site occupation $p$. (b) The wrapping probability $R_2$ vs $p$.
	Vertical line indicates the position of the percolation threshold $p_{\rm c}=0.552\,50(1)$.~\label{fig.kmers8}}
\end{figure}

\begin{figure}[htbp] 
   \centering
        \includegraphics[width=2.8 in]{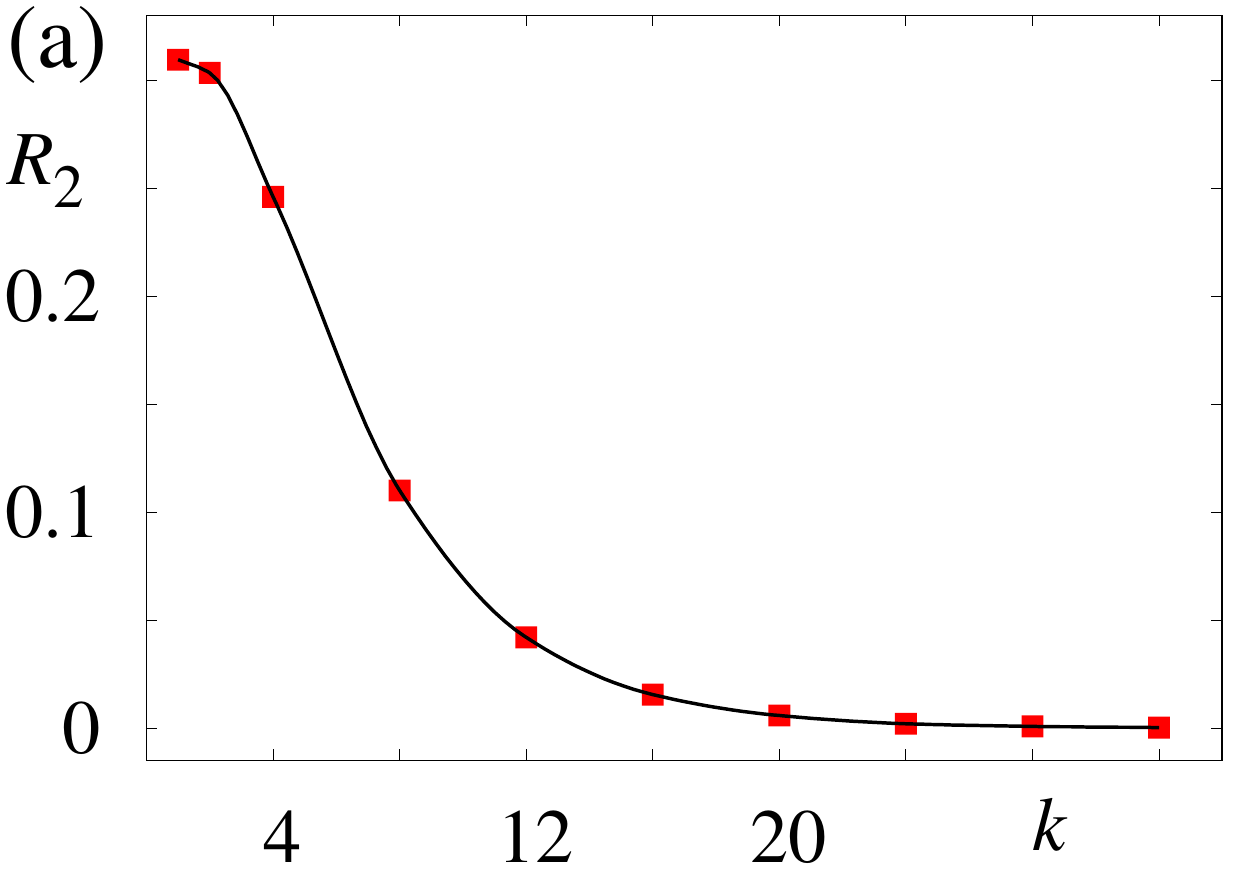}
        \includegraphics[width=2.8 in]{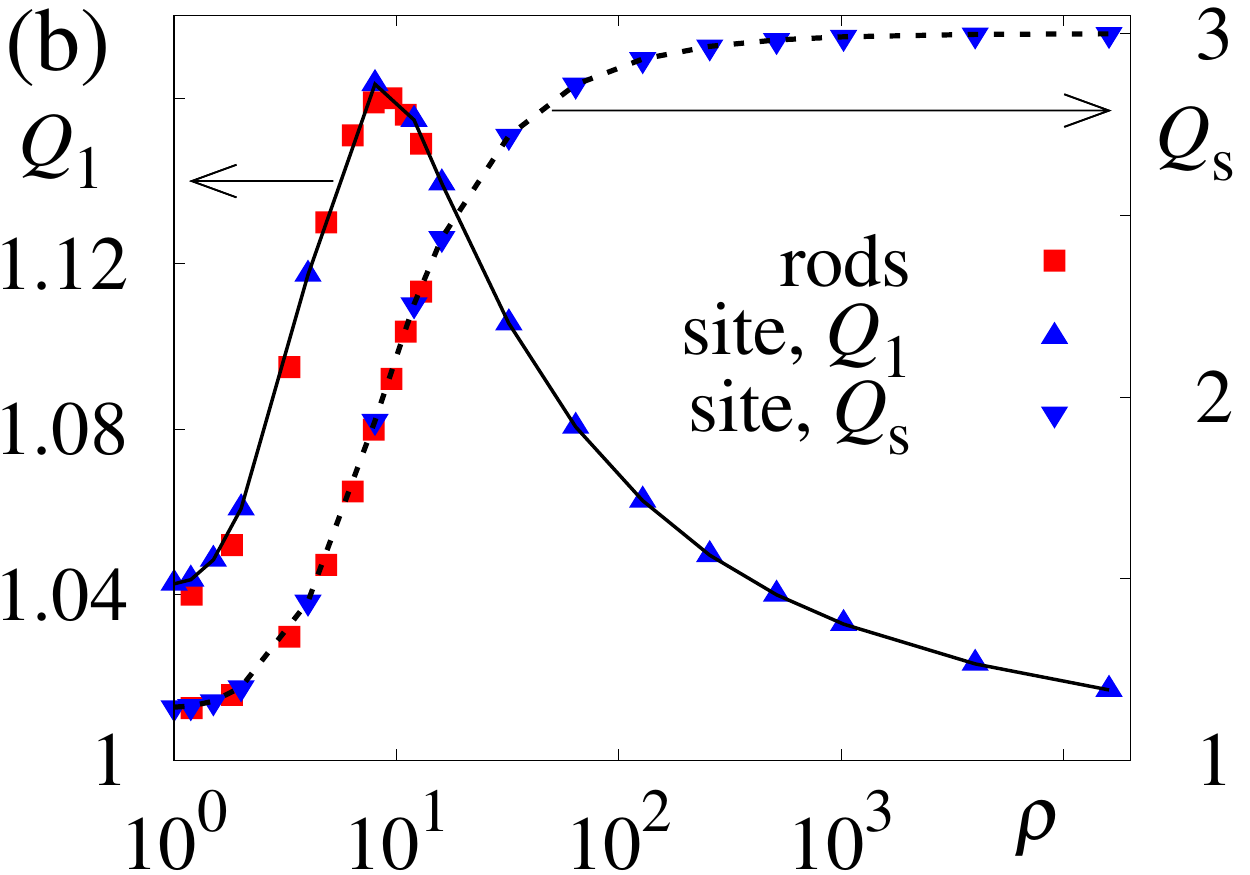}
        \caption{
		Results for percolation of aligned rigid rods on periodic square lattices:
        (a) $R_2$ vs $k$.
        (b) For different $\rho=\rhoe$, values of $Q_1$ and $Q_s$
	are consistent with simulation results for site percolation.}
   \label{fig:kmers}
\end{figure}

 It is seen that, near $p_{\rm c}$, curves of the critical polynomial $\PB$ cross much better than those of wrapping probabilities,
 as exemplified in Fig.~\ref{fig.kmers8}.
 By using the critical polynomial method, we first obtain more precise estimates for the percolation thresholds of aligned 
 rigid rods with different rod sizes $k$, as summarized in Table~\ref{tab.rigidRods}.
 Then, near our estimated thresholds, by finite-size scaling analyses, we determine the critical values of wrapping probabilities and dimensionless ratios. By comparing numerical results for $R_2$, as plotted in Fig.~\ref{fig:kmers}(a), with the theoretical curve for $R_2$ of standard percolation~\cite{Pinson94}, we get values of the effective aspect ratio $\rhoe$ for different $k$ as in Table~\ref{tab.rigidRods}, which is plotted in {Fig.4(a)}. {When $\rhoe=\rho$, from Fig.~\ref{fig:kmers}(b),
it can be seen that critical values of $Q_1$ and $Q_s$ for percolation of the aligned rigid rods are consistent with those for site percolation.}

\begin{table*}[ht]
\begin{center}
\caption{Results for percolation of aligned rigid rods on the square lattice.}
\label{tab.rigidRods}
    \renewcommand\arraystretch{1.25}
    \setlength{\tabcolsep}{3.8mm}{
    \begin{tabular}{lccc}
    \hline
	    $k$ & $p_{\rm c}$ & $R_2^{\rm numer}$ & $\rhoe$ \\
    \hline
	    $2$	& $0.586\,20(5) $ & $0.303\,4(10)$ & $1.20(3)$ \\
	    $4$	& $0.567\,13(3) $ & $0.246(5)$ & $1.82(5)$ \\
	    $8$	& $0.552\,50(1) $ & $0.110(2)$ & $3.30(4)$ \\
	    $12$ & $0.546\,92(2) $ & $0.042(4)$ & $4.83(14)$ \\
	    $16$ & $0.544\,00(4) $ & $0.015\,5(14)$ & $6.36(14)$ \\
	    $20$ & $0.542\,25(3) $ & $0.005\,8(10)$ & $7.9(3)$ \\
	    $24$ & $0.541\,05(2) $ & $0.002\,0(4)$ & $9.5(4)$ \\
	    $28$ & $0.540\,15(3) $ & $0.000\,75(20)$ & $11.0(3)$ \\
	    $32$ & $0.539\,42(4) $ & $0.000\,21(6)$ & $12.9(6)$ \\
    \hline
    \end{tabular}}
\end{center}
\end{table*}

 \section{For anisotropic bond percolation on lattices}
 \label{sec.bond}
 
  \begin{figure}[htbp] 
   \centering
        \includegraphics[width=3.6 in]{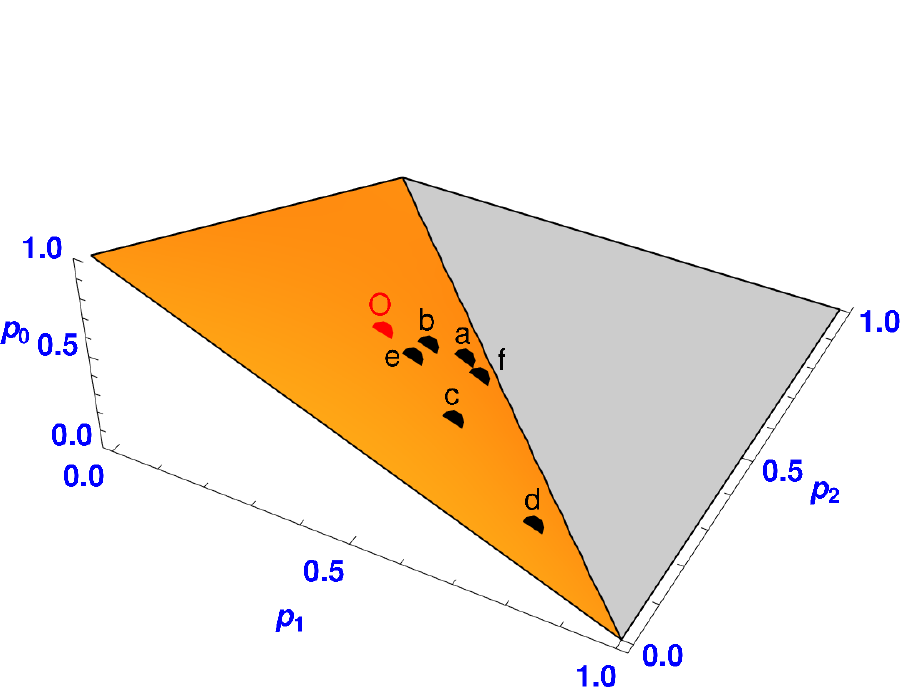}
	  \caption{Selected points on the critical surface $p_0+p_1+p_2-p_0p_1p_2=1$ ($p_i \in [0,1]$)~\cite{SykesEssam1964} for anisotropic bond percolation on the triangular lattice.
        The coordinates of the points (a-f) are given in Table~\ref{Tab:bond-triangle}. Symbol ``O" labels the isotropic point
        with $p_0=p_1=p_2=2\sin{(\pi/18)}$.}
   \label{fig:anisotropicTriPoints}
\end{figure}

 We performed MC simulation for critical bond percolation on $L \times L$ rhombus-shaped periodic triangular lattices,
 with different bond occupation probabilities $(p_0,p_1,p_2)$ (see definition of $p_0$, $p_1$ and $p_2$ in {Fig.~4(b)} of
 the main text) and $L=64,128,256$. The system is symmetric under an arbitrary permutation of the three probabilities.
 Thus we selected points in one-sixth of the critical surface, as shown in Fig.~\ref{fig:anisotropicTriPoints}.
 The number of independent samples for each $(p_0,p_1,p_2,L)$ was $10^6$.
 Using formulae given in the main text, the effective aspect ratio $\rhoe$ and effective boundary twist $t_{\rm e}$ are calculated, as summarized in Table~\ref{Tab:bond-triangle}.
 The resulting numerical values of critical wrapping probabilities are consistent with theoretical predictions, 
 as shown in Table~\ref{Tab:bond-triangle}. 
 
\begin{table*}[ht]
\begin{center}
	\caption{Results for anisotropic bond percolation on the triangular lattice.}
\label{Tab:bond-triangle}
    \renewcommand\arraystretch{1.25}
    \setlength{\tabcolsep}{3.8mm}{
    \begin{tabular}{lccccccc}
    \hline
	    point & $p_0$ & $p_1$ & $p_2$ & $\rhoe$ & $\te$ & $R_2^{\rm theor}$ & $R_2^{\rm numer}$ \\
	    (a) & $1/9$ & $9-\sqrt{73}$ & $9-\sqrt{73}$ & $0.582\,601$ & $1/2$ & $0.310\,556$ & $0.310\,8(9)$ \\
	    (b) & $2/9$ & $(9-\sqrt{67})/2$ & $(9-\sqrt{67})/2$ & $0.692\,306$ & $1/2$ & $0.313\,430$ & $0.314\,0(9)$ \\
	    (c) & $(9-\sqrt{61})/5$ & $5/9$ & $(9-\sqrt{61})/5$ & $0.636\,001$ & $0.228\,312$ & $0.289\,939$ & $0.290\,1(9)$ \\
	    (d) & $(9-\sqrt{67})/7$ & $7/9$ & $(9-\sqrt{67})/7$ & $0.314\,399$ & $0.050\,709$ & $0.123\,794$ & $0.123\,7(6)$ \\
	    (e) & $\sqrt{3}/6$ & $(69-14\sqrt{3})/107$ & $1/3$ &  $0.999\,168$ & $0.453\,880$ & $0.314\,114$ & $0.313\,9(9)$ \\
	    (f) & $2(13-6\sqrt{3})/61$ & $1/2$ & $\sqrt{3}/4$ & $0.558\,158$ & $0.441\,842$ & $0.308\,194$ & $0.308\,5(9)$ \\
    \hline
    \end{tabular}}
\end{center}
\end{table*}

 We also performed MC simulation for critical bond percolation on $L \times L$ periodic square lattices,
 with different perpendicular bond occupation probability $p_{\perp}=p$ (at criticality the parallel 
 bond occupation probability is $1-p_{\perp}$) and $L=1024,2048,4096$.
 The number of independent samples for each $(p_{\perp},L)$ was $O(10^6)$.
 The resulting numerical values of critical wrapping probabilities are also consistent 
 with theoretical predictions, as shown in Table~\ref{Tab:bond-square}.

\begin{figure}[htbp] 
   \centering
        \includegraphics[width=4.0 in]{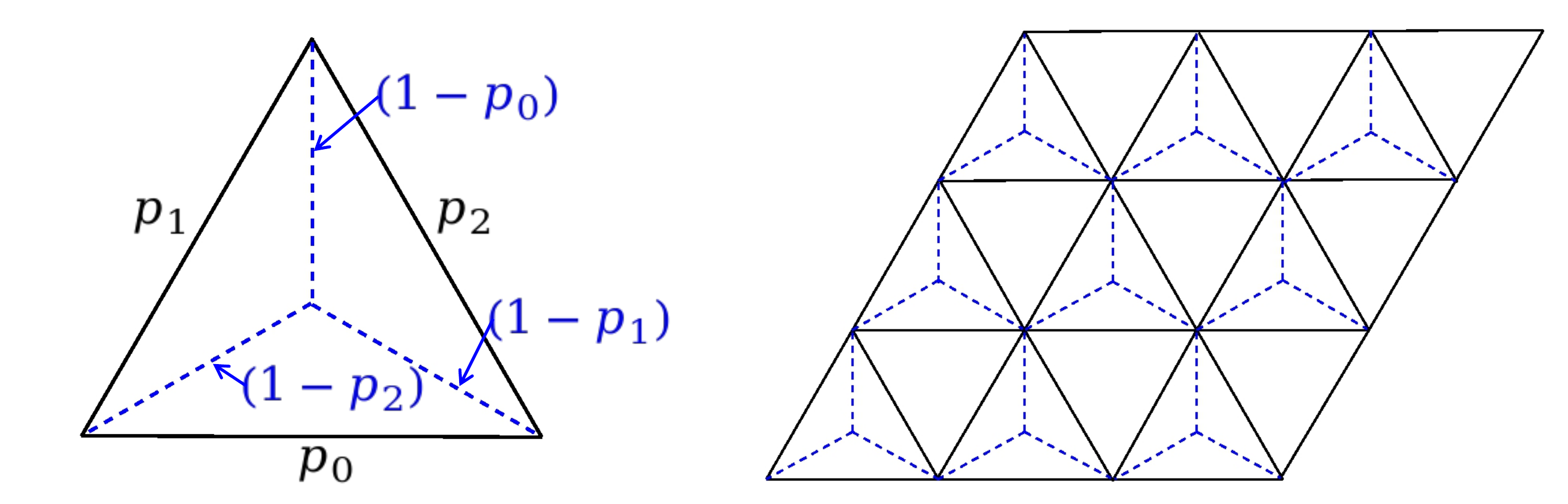}
	\caption{{At criticality, the star-triangle transformation relates anisotropic bond percolation 
	on the honeycomb lattice (dashed lines) to the model on the triangular lattice (solid lines). 
	The relations between critical bond occupation probabilities are given in the left plot, 
	i.e., $p_i^{hon}=1-p_i$ for $i=0,1,2$.}}
   \label{fig:star-triangle}
\end{figure}

 {For anisotropic bond percolation on the honeycomb lattice, the star-triangle transformation relates
 it to the model on the triangular lattice, as shown in Fig.~\ref{fig:star-triangle}. 
 On isoradial graphs under the star-triangle transformation, the two models share the same shape~\cite{GM2014},
 thus $\rhoe$ and $\te$ for the model on the honeycomb lattice with $p_i^{\rm hon}=1-p_i$ ($i=0,1,2$)
 take same values (given in the main text) as the model on the triangular lattice with $(p_0,p_1,p_2)$.}

\begin{table*}[htbp]
\begin{center}
	\caption{Results for anisotropic bond percolation on the square lattice.}
\label{Tab:bond-square}
    \renewcommand\arraystretch{1.25}
    \setlength{\tabcolsep}{3.8mm}{
    \begin{tabular}{lccc}
    \hline
	    $p_{\perp}$ & $\rhoe$ & $R_2^{\rm theor}$ & $R_2^{\rm numer}$ \\
	    $0.1$	& $7.287\,806$ & $0.008\,471$ & $0.008\,3(2)$ \\
	    $0.15$	& $4.708\,181$ & $0.045\,277$ & $0.045\,0(4)$ \\
	    $0.2$	& $3.410\,872$ & $0.102\,558$ & $0.102\,4(6)$ \\ 
	    $0.25$	& $2.626\,051$ & $0.163\,127$ & $0.162\,3(8)$ \\ 
	    $0.3$	& $2.097\,096$ & $0.216\,357$ & $0.216\,2(9)$ \\
	    $0.35$	& $1.714\,016$ & $0.257\,935$ & $0.258\,7(9)$ \\ 
	    $0.4$	& $1.421\,809$ & $0.286\,980$ & $0.286\,3(10)$ \\
	    $0.45$	& $1.189\,916$ & $0.303\,958$ & $0.304\,4(10)$ \\
    \hline
    \end{tabular}}
\end{center}
\end{table*}

 \section{A script for calculating exact wrapping probabilities of standard percolation with given aspect ratio $\rho$ and boundary twist $t$ in two dimensions}
 \label{sec.wrapping}

\begin{figure}[htbp]
\centering
        \includegraphics[width=7.0in]{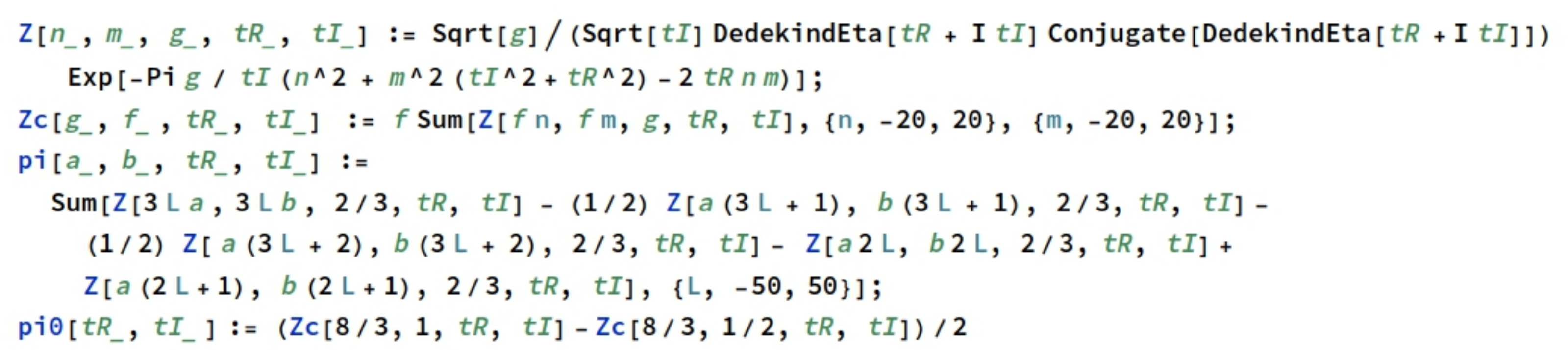}
	\caption{A Mathematica script for calculating exact critical wrapping probabilities of standard percolation in two dimensions.
	Definitions and expressions are given in Ref.~\cite{Pinson94}.
	Putting the script into Mathematica, the exact value of $R_2$ is given by ${\rm pi0}[t,\rho]$.
	The range of integers $n$, $m$ and $L$ taken here are adequate for usual applications,
	and they can be increased to obtain extremely precise values of wrapping probabilities.~\label{fig.script}}
\label{fig:script}
\end{figure}

 Exact values of various critical wrapping probabilities for standard percolation in 2D can be obtained using Eq.~(3.16) in Ref.~\cite{Pinson94},
 e.g., $R_2$ is just $\pi(0)$, whose expression is given in the equation.
 Using variables and functions in Ref.~\cite{Pinson94}, and denoting the aspect ratio $\rho$ and boundary twist $t$ 
 by ``tI" ($\tau_I$, the imaginary part of $\tau$) and ``tR" ($\tau_R$, the real part of $\tau$), respectively, 
 a Mathematica script for Eq.~(1.3), Eq.~(1.5) and Eq.~(3.16) of Ref.~\cite{Pinson94} is presented in Fig.~\ref{fig:script}.

\section{Preliminary results for anisotropic $q$-state Potts model on the triangular lattice}
\label{sec.aniPotts}
For anisotropic $q$-state ($1 \le q \le 4$) Potts model on $L \times L$ periodic triangular lattices, using the isoradial-graph method~\cite{Copin2018}, 
we derive the effective aspect ratio $\rhoe$ and boundary twist $\te$ as
\begin{eqnarray}
        \rhoe = \sin(\frac{\theta_2}{2}) \sin(\frac{\theta_1}{2}) / \sin(\frac{\theta_0}{2}) \, ,
\end{eqnarray}
and 
\begin{eqnarray}
        \te = \cos(\frac{\theta_2}{2}) \sin(\frac{\theta_1}{2}) / \sin(\frac{\theta_0}{2}) \, .
\end{eqnarray}
These two formulas appear same as those for anisotropic bond percolation in the main text, 
but here the angles are given~\cite{Kenyon2004, Copin2018} by
\begin{eqnarray}
        e^{K_i}-1 = \sqrt{q} \frac{\sin(r(\pi-\theta_i))}{\sin(r \theta_i)} \,, \,\, i=0,1,2\,,
\end{eqnarray}
where $K_i$ is the Potts coupling in the $i$-th direction, and $r = \frac{1}{\pi} \cos^{-1}(\frac{\sqrt{q}}{2})$.
For given $\rhoe$ and $\te$, exact wrapping probabilities are predicted by formulas in Ref.~\cite{Arguin2002}. We performed MC simulations by the Swendsen-Wang algorithm for the anisotropic $q=2$, $3$ {and $4$} Potts models on $L \times L$ periodic triangular lattices.
The numerical results are consistent with theoretical predictions, as shown in Table~\ref{Tab:wrapping-q2q3}.
 A full account of results for anisotropic $q$-state Potts models {(including those on the square and honeycomb lattices)} will be published in the future~\cite{HuZiffDeng}.
 
\begin{table*}[ht]
\begin{center}
        \caption{Results for anisotropic $q$-state Potts models on periodic $L \times L$ triangular lattices.
	Numerical results {for $q=2$ and $3$} are from MC simulations with $L=128$.
	{For $q=4$,
        the numerical result is obtained by fitting the MC data to $R_2=R_{2,0}+b/\log{L}$ (here $R_{2,0}$
        represents the value of $R_2$ in the thermodynamic limit, $b$ is a nonuniversal amplitude),
        with $128 \le L \le 2048$.}
	}
\label{Tab:wrapping-q2q3}
    \renewcommand\arraystretch{1.25}
    \setlength{\tabcolsep}{3.8mm}{
    \begin{tabular}{lccccccc}
    \hline
            $q$ & $K_0$ & $K_1$ & $K_2$ & $\rhoe$ & $\te$ & $R_2^{\rm theor}$ & $R_2^{\rm numer}$ \\
            $2$ & $\ln{3}/2$ & $\ln{3}$ & $0.120\,191$ & $0.687\,846$ & $0.082\,872$ & $0.414\,736$ & $0.415(2)$ \\
            $3$ & $0.630\,945$ & $1.261\,889$ & $0.128\,532$ & $0.680\,448$ & $0.075\,916$ & $0.508\,827$ & $0.508(5)$ \\
	    {$4$} & {$\ln{2}$} & {$2 \ln{2}$} & {$0.133\,531$} & {$0.674\,998$} & {$0.070\,945$} & {$0.625\,324$} & {$0.634(27)$} \\
    \hline
    \end{tabular}}
\end{center}
\end{table*}

\bibliographystyle{apsrev4-1}
\bibliography{Janus_revised_SM.bbl}